# How old are dense core vesicles residing in en passant boutons: Simulation of the mean age of dense core vesicles in axonal arbors accounting for resident and transiting vesicle populations

I. A. Kuznetsov[(a), (b)] and A. V. Kuznetsov[(c)]

[(a)] Perelman School of Medicine, University of Pennsylvania, Philadelphia, PA 19104, USA

[(b)] Department of Bioengineering, University of Pennsylvania, Philadelphia, PA 19104, USA

[(c)] Department of MeFchanical and Aerospace Engineering, North Carolina State University, Raleigh, NC 27695-7910, USA; e-mail: avkuznet@ncsu.edu

## Abstract

In neurons, neuropeptides are synthesized in the soma and are then transported along the axon in dense core vesicles (DCVs). DCVs are captured in varicosities located along the axon terminal called *en passant* boutons, which are active terminal sites that accumulate and release neurotransmitters. Recently developed experimental techniques allow for the estimation of the age of DCVs in various locations in the axon terminal. Accurate simulation of the mean age of DCVs in boutons requires the development of a model that would account for resident, transiting-anterograde, and transiting-retrograde DCV populations. In this paper, such a model is developed. The model is applied to simulating DCV transport in *Drosophila* type II motoneurons. The model simulates DCV transport and capture in the axon terminals and makes it possible to predict the age density distribution of DCVs in *en passant* boutons as well as DCV's mean age in boutons. The predicted prevalence of older organelles in distal boutons may explain the "dying back" pattern of axonal degeneration observed in dopaminergic neurons in Parkinson's disease. The predicted difference of two hours between the age of older DCVs residing in distal boutons and the age of younger DCVs residing in proximal boutons is consistent with an approximate estimate of age difference deduced from experimental observations. The age density of resident DCVs is found to be bimodal, which is because DCVs are captured from two transiting states: the anterograde transiting state that contains younger DCVs and the retrograde transiting state that contains older DCVs.

## 1. Introduction

Neuropeptides play an important role in the regulation of mood, motivation, sleep, and drug addiction [1]. Alterations in neuropeptide Y levels may be linked to neurodegenerative and neuroimmune diseases [2]. Neuropeptides are synthesized in the neuron body and then are transported in dense core vesicles (DCVs) through the axon toward the axon terminals (Fig. 1a). Anterograde transport of DCVs is driven by kinesin-1 and kinesin-3 motors while retrograde transport is driven by cytoplasmic dynein motors [3].

*Drosophila melanogaster* is a popular model for investigating DCV transport [4-7]. The above papers investigated DCV transport in axons of *Drosophila* motoneurons with type I, II, and III endings. These axonal endings have different morphologies with different numbers and sizes of *en passant* boutons (hereafter boutons). Boutons are varicosities located along the axon terminals; they are sites that accumulate and release neurotransmitters. Type II terminals have the largest number of boutons in *Drosophila*. The neurons that produce type II boutons in *Drosophila* and the neurons that die in humans with Parkinson's disease (PD) share two properties: they are monoaminergic, which confers sensitivity to autoxidation, and they have extensive axonal arbors [7,8]. Thus a study of neurons with type II terminals may be useful to better understand PD.

Understanding the age distribution of organelles in axons is important because older organelles may have impaired function due to accumulated oxidative damage [9]. Investigating the DCV age distribution may also give clues on how the DCV addressing and delivery system to boutons is designed [5]. For investigating DCV mean age distribution in various boutons of type II terminals, Levitan and colleagues [7] used a photoconvertible construct that switches from green to red fluorescence over a period of hours (red corresponds to older DCVs and green corresponds to younger DCVs). Their results demonstrate that in type II terminals, DCVs residing in distal boutons are older than those residing in proximal boutons.

Models simulating DCV transport in type Ib and type III axon terminals were developed in our previous papers [10-13]. The model of DCV transport in type II terminals, which contain a much larger number of boutons, was developed in [14]. We established that a model that simulates only the resident DCV population cannot correctly predict experimental results of [7] with respect to DCV age (older DCVs in more distal boutons). It was proposed that a model capable of simulating the resident and transiting DCV populations needs to be developed. In the present paper, we develop a model that simulates resident, transiting-anterograde, and transiting-retrograde DCV populations and apply it to the analysis of DCV age distribution in type II terminals.

## 2. Materials and models

### 2.1. Governing equations

We simulated a terminal with 26 boutons (Fig. 1a). The model developed in this paper accounts for DCV concentrations in two different types of kinetic states in each bouton, the transiting and resident states. In the resident states DCVs stay captured in boutons while in the transiting states DCVs can move between the boutons (Fig. 1b). The transiting states are further divided into anterograde and retrograde (Fig. 2). This is done for all transiting states except in bouton 1 (because DCVs turn around in bouton 1, this transiting state in neither anterograde nor retrograde). Our previous model, developed in [14], accounts only for DCVs in the resident states and contains 26 ordinary differential equations (ODEs). The new model accounts for 26 resident and 51 transiting kinetic states and contains 77 ODEs. There are coupling terms between the transiting and resident DCVs in the new model (a transiting DCV can be captured and become a resident DCV, and then it can be re-released and become a transiting DCV again). Therefore, we need to state the governing equations for the new model.

We numbered the boutons in the axon terminal from the most distal #1 to the most proximal #26, following the convention introduced in [5] (Fig. 1a). Since the length of an axon is much greater than its width, we used the linear number density to characterize the DCV concentration. We thus defined the DCV concentration as the number of DCVs (resident or transiting) per unit length of the axon.

We used a multi-compartment model [15-17] to formulate equations expressing the conservation of the number of DCVs in the compartments. Most boutons have three compartments (corresponding to resident, transiting-anterograde, and transiting-retrograde states) while bouton 1 has two compartments (corresponding to resident and transiting states), see Fig. 2. The number of DCVs is the conserved quantity.

The conservation of resident DCVs in the most proximal bouton gives the following equation (Fig. 2):

$$L_{26} \frac{dn_{26}}{dt} = \min\left[ h_{26}^a \left( n_{sat0,26} - n_{26} \right), j_{ax \to 26} \right] + \min\left[ h_{26}^r \left( n_{sat0,26} - n_{26} \right), j_{25 \to 26} \right] - L_{26} \frac{n_{26} \ln(2)}{T_{1/2}}, \quad (1)$$

where $n_i$ is the concentration of resident DCVs in bouton $i$ ($i=1,\ldots,26$); $t$ is the time; $n_{sat0,i}$ is the saturated concentration of resident DCVs in bouton $i$ at infinite DCV half-life or at infinite DCV residence time ($i=1,\ldots,26$); $L_i$ is the length of a compartment occupied by bouton $i$ (defined in Fig. 1a, $i=1,\ldots,26$); $h_i^a$ and $h_i^r$ are the mass transfer coefficients characterizing the rates of DCVs capture into the resident state in bouton $i$ ($2,\ldots,26$) as DCVs pass through bouton $i$ in the anterograde and retrograde directions, respectively; $j_{ax\to26}$ is the anterograde flux of new DCVs from the axon to the most proximal bouton (#26); and $T_{1/2}$ is the half-life or half-residence time of DCVs captured into the resident state in boutons. In terms of DCV conservation in the resident states in boutons, it does not matter whether DCVs (and their content) are destroyed or released back to the transiting pool. If DCVs are destroyed (this case also includes the situation of DCV consumption in boutons due to neuropeptides release by exocytosis), $T_{1/2}$ is interpreted as the half-life. If DCVs are released back to circulation, $T_{1/2}$ is interpreted as the half-residence time. At a first approximation, we assume that processes such as DCV destruction in boutons, DCV consumption due to neuropeptide exocytosis, and DCVs release back to circulation, are described by linear kinetics, and hence, can be modeled by a term similar to the last term on the right-hand side of Eq. (1) (the term involving $T_{1/2}$).

The term $\min\left[h_{26}^a\left(n_{sat0,26}-n_{26}\right),j_{ax\to26}\right]$ on the right-hand side of Eq. (1) simulates the fact that the rate at which a bouton captures anterogradely moving DCVs cannot exceed the flux of anterogradely moving DCVs into the bouton. A similar role is played by the term $\min\left[h_{26}^r\left(n_{sat0,26}-n_{26}\right),j_{25\to26}\right]$.

Stating the conservation of anterograde transiting DCVs in the most proximal bouton results in the following equation (Fig. 2):

$$L_{26}\frac{dn_{26,t}^a}{dt}=j_{ax\to26}-j_{26\to25}-\min\left[h_{26}^a\left(n_{sat0,26}-n_{26}\right),j_{ax\to26}\right]+\varepsilon\delta L_{26}\frac{n_{26}\ln(2)}{T_{1/2}}, \qquad (2)$$

where $n_{26,t}^a$ is the concentration of anterograde transiting DCVs in bouton $i$ ($i=2,\ldots,26$), $j_{j\to k}$ is the flux of DCVs from compartment "$j$" to compartment "$k$" (see Fig. 1b), $j_{26\to ax}$ is the retrograde flux of DCVs from the most proximal bouton back to the axon, and $\delta$ is the parameter which determines what happens to DCVs captured into the resident state of the boutons. $\delta=0$ means that all DCVs are eventually destroyed in boutons (or the DCV content leaves the boutons by spontaneous exocytotic events) while $\delta = 1$ means that DCVs, after spending some time in the resident state, are released back to the transiting pool.

In Eq. (2) $\varepsilon$ is the parameter indicating how DCVs released from the resident state are split between the anterograde and retrograde components. DCVs released from the resident state first join the transiting state in the corresponding bouton. $\varepsilon$ is the portion of released DCVs that join the anterograde component and $(1-\varepsilon)$ is the portion of DCVs that join the retrograde component.

The term release is used in our model as a term opposing capture. In biological literature, release also refers to releasing the contents inside of vesicles to the extracellular milieu by exocytosis. This is different from release of a captured DCV back to the transiting pool, which is the meaning of the term release in the context of our model.

Stating the conservation of retrograde transiting DCVs in the most proximal bouton gives the following equation (Fig. 2):

$$L_{26} \frac{dn_{26,t}^r}{dt} = j_{25 \rightarrow 26} - j_{26 \rightarrow ax} - \min\left[ h_{26}^r \left( n_{sat0,26} - n_{26} \right), j_{25 \rightarrow 26} \right] + (1-\varepsilon)\delta L_{26} \frac{n_{26} \ln(2)}{T_{1/2}}, \qquad (3)$$

where $n_{26,t}^r$ is the concentration of retrograde transiting DCVs in bouton $i$ ($i$=2,…,26).

In boutons 25 through 2, stating the conservation of resident DCVs gives the following equations (Fig. 2):

$$L_i \frac{dn_i}{dt} = \min\left[ h_i^a \left( n_{sat0,i} - n_i \right), j_{i+1 \rightarrow i} \right] + \min\left[ h_i^r \left( n_{sat0,i} - n_i \right), j_{i-1 \rightarrow i} \right] - L_i \frac{n_i \ln(2)}{T_{1/2}} \qquad (i=25,…,2).$$

$$(4)$$

Stating the conservation of anterograde transiting DCVs in boutons 25 through 2 gives the following equations (Fig. 2):

$$L_i \frac{dn_{i,t}^a}{dt} = j_{i+1 \rightarrow i} - j_{i \rightarrow i-1} - \min\left[ h_i^a \left( n_{sat0,i} - n_i \right), j_{i+1 \rightarrow i} \right] + \varepsilon\delta L_i \frac{n_i \ln(2)}{T_{1/2}} \qquad (i=25,…,2).$$

$$(5)$$

Also, stating the conservation of retrograde transiting DCVs in boutons 25 through 2 gives the following equations (Fig. 2):

$$L_i \frac{dn_{i,t}^r}{dt} = j_{i-1 \rightarrow i} - j_{i \rightarrow i+1} - \min\left[ h_i^r \left( n_{sat0,i} - n_i \right), j_{i-1 \rightarrow i} \right] + \left(1-\varepsilon\right) \delta L_i \frac{n_i \ln(2)}{T_{1/2}} \quad (i=25,\ldots,2).$$

(6)

Stating the conservation of resident DCVs in the most distal bouton gives the following equation (Fig. 2):

$$L_1 \frac{dn_1}{dt} = \min\left[ h_1 \left( n_{sat0,1} - n_1 \right), j_{2 \rightarrow 1} \right] - L_1 \frac{n_1 \ln(2)}{T_{1/2}},$$

(7)

where $h_1$ is the mass transfer coefficient characterizing the rate of DCVs capture into the resident state in bouton 1 (DCVs pass bouton 1 only once).

There are no separate anterograde and retrograde transiting states in bouton 1 because DCVs turn around in this bouton. The conservation of transiting DCVs in the most distal bouton is used to produce the following equation (Fig. 2):

$$L_1 \frac{dn_{1,t}}{dt} = j_{2 \rightarrow 1} - j_{1 \rightarrow 2} - \min\left[ h_1 \left( n_{sat0,1} - n_1 \right), j_{2 \rightarrow 1} \right] + \delta L_1 \frac{n_1 \ln(2)}{T_{1/2}},$$

(8)

where $n_{1,t}$ is the concentration of transiting DCVs in bouton 1. In this paper we assumed that $j_{ax \rightarrow 26}$ remains constant during the process of filling the terminal. Modeling the time dependence of $j_{ax \rightarrow 26}$ would require simulating the DCV concentration in the axon, which can be done as described in section S1 of Supplemental Materials.

Eqs. (1)-(8) include fluxes in the terminal (Fig. 1b), which need to be modeled. The DCV fluxes have units of vesicles/s. The anterograde flux of transiting DCVs between the most proximal bouton (bouton 26) and bouton 25 is simulated as follows:

$$j_{26 \rightarrow 25} = j_{ax \rightarrow 26} - \min\left[ h_{26}^a \left( n_{sat0,26} - n_{26} \right), j_{ax \rightarrow 26} \right] + \varepsilon \delta L_{26} \frac{n_{26} \ln(2)}{T_{1/2}}.$$

(9)

We then used the following equations to model anterograde fluxes between boutons 25 through 2:

$$j_{i \rightarrow i-1} = j_{i+1 \rightarrow i} - \min\left[ h_i^a \left( n_{sat0,i} - n_i \right), j_{i+1 \rightarrow i} \right] + \varepsilon \delta L_i \frac{n_i \ln(2)}{T_{1/2}} \qquad (i=25,\ldots,2).$$

(10)

We modeled the retrograde flux from bouton 1 into bouton 2 by the following equation:

$$j_{1\to2} = H\left[t-t_1\right]\left\{j_{2\to1} - \min\left[h_1\left(n_{sat0,1}-n_1\right), j_{2\to1}\right]\right\} + \delta L_1 \frac{n_1 \ln(2)}{T_{1/2}}, \tag{11}$$

where $H$ is the Heaviside step function and $t_1$ is the time required for DCVs to change the direction in the most distal bouton if they are not captured. The Heaviside step function thus delays the onset of retrograde DCV flux by the time $t_1$. The effect of this initial delay is expected to be small because $t_1 = 300\text{s}$ (Table S1), which is small compared with how long it takes to fill the boutons with DCVs (hours), see Table 1.

We modeled retrograde fluxes between boutons 2 through 25 by using the following equations:

$$j_{i\to i+1} = H\left[t-t_1\right]\left\{j_{i-1\to i} - \min\left[h_i^r\left(n_{sat0,i}-n_i\right), j_{i-1\to i}\right]\right\} + (1-\varepsilon)\delta L_i \frac{n_i \ln(2)}{T_{1/2}} \qquad (i=2,\ldots,25).$$

$$\tag{12}$$

The retrograde flux leaving the terminal from the most proximal bouton is modeled as:

$$j_{26\to ax} = H\left[t-t_1\right]\left\{j_{25\to26} - \min\left[h_{26}^r\left(n_{sat0,26}-n_{26}\right), j_{25\to26}\right]\right\} + (1-\varepsilon)\delta L_{26} \frac{n_{26} \ln(2)}{T_{1/2}}. \tag{13}$$

Eqs. (1)-(8) describe a system of 77 first-order ordinary differential equations (ODEs); 77 initial conditions are thus required.

We assumed that initially there are no DCVs in the resident state in the terminal:

$$n_1(0) = 0, \ldots, n_{26}(0) = 0. \tag{14}$$

We also assumed that the initial DCV concentration in the transiting states is constant and uniform:

$$n_{1,t}(0) = n_{0,t}, n_{2,t}^a(0) = n_{0,t}, n_{2,t}^r(0) = n_{0,t} \ldots, n_{26,t}^a(0) = n_{0,t}, n_{26,t}^r(0) = n_{0,t}. \tag{15}$$

We investigated the sensitivity of the solution to various values of $n_{0,t}$. Values of parameters involved in the model are estimated in section S2 of Supplemental Materials.

## 2.2. Model of age distribution of DCVs and mean age of DCVs in boutons

We used the method developed in [18,19] to compute the age distributions in the resident and transiting states in boutons. We recast governing Eqs. (1)-(8) into the following matrix form:

$$\frac{d}{dt}\mathbf{N}^*(t) = \mathrm{B}\left(\mathbf{N}^*(t),t\right)\mathbf{N}^*(t) + \mathbf{u}(t),\tag{16}$$

where $\mathbf{N}^*$ is the state vector defined by the following components:

$$N_1^* = n_1 L_1, \ldots, N_{26}^* = n_{26} L_{26}, \ N_{27}^* = n_{1,t} L_1, \ N_{28}^* = n_{2,t}^a L_2, \ldots, N_{52}^* = n_{26,t}^a L_{26}, \ N_{53}^* = n_{2,t}^r L_2, \ N_{77}^* = n_{26,t}^r L_{26}.\tag{17}$$

The first 26 components of the state vector represent the number of DCVs in the resident states and the last 51 components represent the number of DCVs in the transiting states (see Fig. 2 and section S3). Vector $\mathbf{u}(t)$ is defined in Eqs. (41) and (42) below.

Matrix B(77,77) in our case is as follows. In simulating DCV fluxes in the most distal bouton (the left-hand side diagram in Fig. 2), we accounted for the internal fluxes between the compartments, the external DCV flux entering the terminal from the axon, the DCV flux leaving the terminal back to the axon, as well as possible destruction of DCVs in the resident state in boutons. Based on the analysis of the DCV fluxes to and from the resident and transiting states in the most distal bouton, equations for the following elements of matrix B were obtained:

$$b_{26,26} = -L_{26}\frac{n_{26}\ln(2)}{T_{1/2}} / \left(L_{26}n_{26}\right) = -\frac{\ln(2)}{T_{1/2}},\tag{18}$$

$$b_{52,26} = \varepsilon\delta\frac{\ln(2)}{T_{1/2}},\tag{19}$$

$$b_{77,26} = (1-\varepsilon)\delta\frac{\ln(2)}{T_{1/2}}.\tag{20}$$

$$b_{52,52} = -[\min\left[h_{26}^a\left(n_{sat0,26} - n_{26}\right), j_{ax\to 26}\right] + j_{26\to 25}] / \left(L_{26}n_{26,t}^a\right),\tag{21}$$

$$b_{26,52} = [\min\left[h_{26}^a\left(n_{sat0,26} - n_{26}\right), j_{ax\to 26}\right]] / \left(L_{26}n_{26,t}^a\right),\tag{22}$$

$$b_{77,76} = j_{25\to 26} / \left(L_{25}n_{25,t}^r\right),\tag{23}$$

$$b_{26,77} = [\min\left[h_{26}^r\left(n_{sat0,26} - n_{26}\right), j_{25\to 26}\right]] / \left(L_{26}n_{26,t}^r\right)\tag{24}$$

$$b_{77,77} = -[\min\left[h_{26}^r\left(n_{sat0,26} - n_{26}\right), j_{25\to 26}\right] + j_{26\to ax}] / \left(L_{26}n_{26,t}^r\right)\tag{25}$$

By repeating a similar analysis for bouton $i$ ($i=2,...,25$) (the middle diagram in Fig. 2), equations for the following elements of matrix B were obtained:

$$b_{i,i} = -L_i \frac{n_i \ln(2)}{T_{1/2}} / \left(L_i n_i\right) = -\frac{\ln(2)}{T_{1/2}},$$
(26)

$$b_{i+26,i} = \varepsilon \delta \frac{\ln(2)}{T_{1/2}}$$
(27)

$$b_{i+51,i} = (1-\varepsilon)\delta \frac{\ln(2)}{T_{1/2}}.$$
(28)

$$b_{i+26,i+26} = -[\min\left[h_i^a\left(n_{sat0,i}-n_i\right), j_{i+1\rightarrow i}\right] + j_{i\rightarrow i-1}] / \left(L_i n_{i,t}^a\right),$$
(29)

$$b_{i,i+26} = [\min\left[h_i^a\left(n_{sat0,i}-n_i\right), j_{i+1\rightarrow i}\right]] / \left(L_i n_{i,t}^a\right),$$
(30)

$$b_{i,i+51} = [\min\left[h_i^r\left(n_{sat0,i}-n_i\right), j_{i-1\rightarrow i}\right]] / \left(L_i n_{i,t}^r\right),$$
(31)

$$b_{i+51,i+51} = -[\min\left[h_i^r\left(n_{sat0,i}-n_i\right), j_{i-1\rightarrow i}\right] + j_{i\rightarrow i+1}] / \left(L_i n_{i,t}^r\right),$$
(32)

$$b_{i+26,i+1+26} = j_{i+1\rightarrow i} / \left(L_{i+1} n_{i+1,t}^a\right),$$
(33)

Additionally, the element $b_{53,27}$ is assigned the following value (note that the flux $j_{1\rightarrow 2}$ leaves the compartment that simulates the transiting state in bouton 1 (with DCV concentration $n_{1,t}$), and enters the compartment that simulates the transiting-retrograde state in bouton 2 (with DCV concentration $n_{2,t}^r$), see Fig. 2):

$$b_{53,27} = j_{1\rightarrow 2} / \left(L_1 n_{1,t}\right),$$
(34)

For $i = 3,...,25$ the element $b_{i+51,i-1+51}$ is assigned the following value:

$$b_{i+51,i-1+51} = j_{i-1\rightarrow i} / \left(L_{i-1} n_{i-1,t}^r\right),$$
(35)

Finally, by repeating a similar analysis for the most distal bouton (the right-hand side diagram in Fig. 2), the following equations were obtained:

$$b_{1,1} = -L_1 \frac{n_1 \ln(2)}{T_{1/2}} / \left(L_1 n_1\right) = -\frac{\ln(2)}{T_{1/2}},$$
(36)

$$b_{27,1} = \delta \frac{\ln(2)}{T_{1/2}} \,. \tag{37}$$

$$b_{1,27} = [\min[h_1(n_{sat0,1} - n_1), j_{2\to1}]] / (L_1 n_{1,t}), \tag{38}$$

$$b_{27,27} = -[\min[h_1(n_{sat0,1} - n_1), j_{2\to1}] + j_{1\to2}] / (L_1 n_{1,t}), \tag{39}$$

$$b_{27,28} = j_{2\to1} / (L_2 n_{2,t}^a), \tag{40}$$

All other elements of matrix B are equal to zero.

The only flux entering the terminal is the anterograde flux from the axon to the most proximal bouton, $j_{ax\to26}$. We assumed that all DCVs which leave the terminal (their flux is described by $j_{26\to ax}$) return to the soma for degradation, and that none of them reenter to the terminal (allowing some DCVs to reverse direction near the soma, as has been seen in motor neurons [5], would shift the age distribution of DCVs in boutons towards an older age, but would not qualitatively affect the results). Thus, the DCVs that enter the terminal (their flux is described by $j_{ax\to26}$) are all newly synthesized in the soma, and their age at the time of entry to the terminal is set to zero. This means that our simulations neglect the time that it takes for DCVs to transit from the soma to the terminal. Thus, the DCV age computed here should be interpreted as the age of DCVs since their entry into the terminal. The 52nd element of vector **u** is then given by the following equation:

$$u_{52} = j_{ax\to26} \,. \tag{41}$$

Since no other external fluxes enter the terminal,

$$u_i = 0 \quad (i=1,\dots,51,53,\dots,77). \tag{42}$$

According to [18], the state transition matrix, $\Phi$, can be found by solving the following matrix equation:

$$\frac{d}{dt}\Phi(t,t_0) = B(\mathbf{N}^*(t),t)\Phi(t,t_0) \,. \tag{43}$$

Eq. (43) must be solved subject to the following initial condition:

$$\Phi(t_0,t_0) = I \,, \tag{44}$$

where I denotes an identity matrix.

We assumed that initially the resident states in boutons do not contain any DCVs and that all DCVs in the transiting states are new. We also assumed that all DCVs entering the terminal are new. Then the age density of DCVs that entered the terminal after $t = 0$ can be calculated as:

$$\mathbf{p}(c,t) = 1_{[0,t-t_0)}(c)\Phi(t,t-c)\mathbf{u}(t-c),\tag{45}$$

where $1_{[0,t-t_0)}$ is the indicator function which is equal to 1 if $0 \le c < t - t_0$, otherwise, $1_{[0,t-t_0)}$ equals 0. The age density of DCVs can be understood as the ratio of the number of DCVs having an age between $T$ and $T+dT$ over the duration of this interval $dT$. The integral from 0 to infinity with respect to time of the age density of DCVs in a certain kinetic state gives the number of DCVs in this kinetic state. More generally, an integral over any time period gives the number of DCVs having an age within that time range.

Following [20], the mean age of DCVs in boutons is defined as:

$$\bar{c}_i(t) = \frac{\int\limits_0^\infty c p_i(c,t)\,dc}{\int\limits_0^\infty p_i(c,t)\,dc} \qquad (i=1,\ldots,77),\tag{46}$$

where $1 \le i \le 26$ corresponds to the resident states in boutons and $27 \le i \le 77$ corresponds to the transiting states in boutons. According to [18,20], $\bar{c}_i(t)$ $(i=1,\ldots,77)$ can be found by solving the following mean age system:

$$\frac{d}{dt}\bar{\mathbf{c}}(t) = \mathbf{g}\left(\mathbf{N}^*(t),t\right),\tag{47}$$

which must be solved subject to the following initial condition:

$$\bar{\mathbf{c}}(0) = 0,\tag{48}$$

where $\bar{\mathbf{c}} = (\bar{c}_1,\ldots,\bar{c}_{77})$ and $\mathbf{g}$ is a vector defined as follows:

$$g_i(t) = 1 + \frac{\sum\limits_{j=1}^{77}\left(\bar{c}_j(t) - \bar{c}_i(t)\right)b_{i,j}(t)N_j(t) - \bar{c}_i(t)u_i}{N_i(t)} \qquad (i=1,\ldots,77).\tag{49}$$

We then defined the following vectors that characterize the mean age of DCVs in the resident and transiting states in boutons, respectively:

$$\bar{\mathbf{a}} = (\bar{c}_1,\ldots,\bar{c}_{26}),\ \bar{\mathbf{a}}_t = (\bar{c}_{27},\ldots,\bar{c}_{77}),\tag{50}$$

where $\overline{\mathbf{a}}$ and $\overline{\mathbf{a}}_l$ are one-dimensional arrays of size 26 and 51, respectively.

Details of the numerical solution are given in section S4.

## 3. Results

### 3.1. Assumptions concerning the fate of DCVs captured into the resident state in boutons

In this paper, we investigate two possible scenarios with respect to DCV destruction (consumption) in boutons. (i) $\delta = 1$. This scenario simulates the situation when all captured DCVs, after spending some time in the resident state, are re-released back to the transiting state. This scenario is supported by the scarcity of the organelle degradation machinery in axons [21], which suggests that instead of being destroyed in boutons, captured DCVs may be released back to the transiting pool and re-enter the circulation. Old DCVs may then exit the circulation and travel back to the soma, where they may be destroyed in lysosomes, which are plentiful in the soma. In this scenario, since older DCVs are returning to the transiting state, the average age of DCVs is expected to increase faster from the proximal to distal boutons than in the second scenario. The increase of the DCV age from proximal to distal boutons is supported by experimental observations reported in [7]. The results for $\delta = 1$ are reported in Figs. 3-6 and S4-S7. It should be noted that the results for the DCV concentrations in the resident states in boutons, shown in Figs. 3 and 4 and in Table 1, are almost independent of the value of $\delta$ because it does not matter how DCVs leave the resident state, by re-release of DCVs back to the transiting state or by DCV destruction. The value of $\delta$ has some effect at small times because capture into the resident state is limited by the number of DCVs that enter the boutons, and DCV fluxes between the boutons are larger for $\delta = 1$ than for $\delta = 0$. (ii) $\delta = 0$. This scenario simulates the situation in which all captured DCVs are eventually destroyed or consumed in boutons. This scenario is supported by the following argument. Neuropeptides released from DCVs may be marked by ubiquitination for degradation in proteasomes. Also, neuropeptides may be released from boutons by spontaneous secretory events, which will also lead to DCV consumption in boutons. The results for $\delta = 0$ are reported in Figs. S2 and S3.

### 3.2. Verifying values to which concentrations of resident DCVs converge as $t \rightarrow \infty$

The solution of Eqs. (1)-(8) was verified by comparing DCV concentrations in the resident states at $t \rightarrow \infty$ (at steady-state) with the estimates of these parameters given by Eqs. (S4) and (S5). The concentrations converge to correct estimates of steady-state values (Fig. 3a). Also, Fig. 3a shows that we were able to incorporate into our model the drop-off in the DCV content (reported in [7]) that appeared suddenly at the farthest ends of the arbor. The model suggests that the drop-off could be explained by the assumption that approximately 20% of the most distal boutons are characterized by different parameter values. According to our hypothesis, these boutons have more limited capacity than other boutons, compare Eqs. (S4) and (S5).

Another possible explanation is that the drop-off is caused by a depletion of the anterograde flux so that the most distal boutons do not receive enough DCVs. The fact that the drop-off distally is so sudden (Fig. 3a) may be indicative of an instability caused by an imbalance between anterograde and retrograde DCV fluxes. This hypothesis is supported by an observation that the drop-off does not occur in all branches, but rather is more common in the most distal branches [7].

More experimental research is needed in order to understand whether the position of the drop-off advances over time or remains between the same boutons (in our case, between boutons 5 and 4). If the latter is true, the explanation of the drop-off by limited capacity of the farthest boutons is plausible. An interesting physiological question is whether the drop-off means that the most distal boutons remain unused for neuropeptide release.

We assumed that the rate at which DCVs enter an anterograde transiting state of a bouton equals the rate at which DCVs leave the anterograde transiting state, see Eqs. (2) and (5) and Fig. 2. The same is assumed about the retrograde transiting state, see Eqs. (3) and (6) and Fig. 2. Also, the same is assumed about the transiting state in bouton 1, see Eq. (8) and Fig. (2). Therefore, the DCV concentrations in the transiting states are equal to their initial concentrations, $n_{0,t}$, over the whole duration of the process of filling the terminal with DCVs. Values of $n_{0,t}$, postulated in Eq. (15), are the same for all boutons (data not shown).

To check the order in which the boutons are filled, we plotted concentrations of resident DCVs in boutons at three representative times ($t$ = 1, 2, and 5 h). The results reported in Fig. 3b agree with [7] which reported that in type II terminals DCVs first accumulate in proximal boutons; accumulation in distal boutons occurs much slower (and later) than in proximal boutons. It is interesting that the modeling results show that this behavior is more pronounced later in the process. A dip in the curve for $t$ = 1 h, for example, suggests that in the beginning some boutons (for example, bouton 19) accumulate less DCVs than more distal boutons (Fig. 3b).

### 3.3. The buildup of DCV concentrations in the resident and transiting states in boutons

It takes less than 40 hours for the DCV concentrations in the resident state in boutons to reach their steady-state values. Interestingly, bouton 5 takes the longest amount of time to fill while bouton 1 takes the shortest amount of time (Fig. 4, Table 1). This is because we explained the drop-off in the DCV content, reported in [7], by assuming a smaller DCV capacity of the four most distal boutons (Eq. (S5)). Since capacity of boutons 1-4 is assumed to be small, it takes a short time to fill them. Bouton 5 is the most distal bouton with a large capacity, and since the anterograde flux is reduced with distance by capture in more proximal boutons, it takes the longest time to fill bouton 5. The time to reach steady-state concentration in the resident state does not depend on the value of $n_{0,t}$ and only slightly depends on the value of $\varepsilon$ (Table 1).

Table 1. Approximate time, $t_\infty$, required for resident DCVs to reach a steady-state concentration in four representative boutons (#1, 5, 13, and 26). We assumed that steady-state is reached when the DCV concentration in a particular bouton reaches 99% of $n_{sat,i}$, which is defined in Eqs. (S4) and (S5). Since the DCV fluxes in and out the transiting states are assumed to be the same, concentrations of transiting DCVs stay constant (equal to the value postulated by Eq. (15)) throughout the process of filling the terminal. It should be noted that the duration of *Drosophila* third instar (used in experiments of [7]) is approximately 48 hours. $\delta = 1$.

| $n_{0,t}$ | $\varepsilon$ | $t_\infty$ for $n_1$ (h) | $t_\infty$ for $n_5$ (h) | $t_\infty$ for $n_{13}$ (h) | $t_\infty$ for $n_{26}$ (h) |
|---|---|---|---|---|---|
| 1 | 0.8 | 1.82 | 39.42 | 19.99 | 7.58 |
| 0.1 | 0.5 | 1.82 | 39.42 | 19.99 | 7.58 |
| 1 | 0 | 1.82 | 39.42 | 20.02 | 7.49 |
| 1 | 1 | 1.82 | 39.42 | 19.98 | 7.60 |

### 3.4. DCV age density distributions and mean DCV ages for the case when all captured DCVs are re-released to the circulation, $\delta = 1$

At steady-state, the age of resident vesicles is distributed between 0 and 60 hours. The age density of resident DCVs is bimodal, which occurs because DCVs are captured into the resident state from anterograde (younger) and retrograde (older) transiting pools (Fig. 2).

The age density of transiting DCVs exhibits a peak at a certain age, which shifts toward an older age from more proximal to more distal boutons for anterograde transiting DCVs (Fig. 5b) and from more distal to more proximal boutons for retrograde transiting DCVs (Fig. 5c). The shift of the peak density thus occurs in the direction in which DCVs travel, as older DCVs travel to the next bouton.

Intriguingly, at steady-state the peak on the age density curve for DCVs in the transiting states does not coincide with the mean ages of DCVs in the transiting states (compare Figs. 5b,c and 6b,c). For example, the peak on the curve showing the age density of DCVs in the retrograde transiting state in bouton 2 (see the line marked by hollow circles in Fig. 5c) occurs at the DCV age of 1.94 hours, while the mean DCV age in the retrograde kinetic state in bouton 2 is 6.88 hours (Fig. 6c).

This is explained by the fact that although the age density distributions of transiting DCVs in Figs. 5b,c look similar to bell-shaped curves resembling normal distributions, they are in fact rightward skewed. Note that the age density in Figs. 5b,c takes a small positive value even at large values of $t$, which shifts the mean of the age density distribution to the right. For example, for $\delta = 1$, which corresponds to the situation when DCVs captured into the resident state are eventually released back to the circulation (the case displayed in Figs. 5 and 6), the age density of transiting-retrograde DCVs in bouton 2 at 6 hours equals to 0.2100 vesicle/($\mu$m h). This small positive value corresponds to older DCVs released from the resident state in boutons. Note that this effect is absent if $\delta = 0$, which corresponds to the situation when DCVs captured into the resident state are eventually destroyed in boutons or leave the boutons by exocytosis. This situation is displayed in Figs. S2b,c. Indeed, the age density of transiting-retrograde DCVs in bouton 2 at 6 hours is equal to 0.0001 vesicle/($\mu$m h) (see the line marked by hollow circles in Fig. S2c). The peaks on the age density distributions in Figs. S2b,c exactly correspond to the mean ages of DCVs displayed in Figs. S3b,c.

The accuracy of the computed age density distributions was checked by integrating the densities with respect to the DCV age in boutons (which should give the number of DCVs) and comparing the result with the number of DCVs computed by solving Eqs. (1)-(8) with boundary conditions (14), (15) (Fig. S1). The number of DCVs is equal to the DCV concentration (found by solving Eqs. (1)-(8)) multiplied by the length of a compartment.

The mean age of resident DCVs changes from 13.55 hours in the most proximal bouton (#26) to 15.46 hours in the most distal bouton (#1) (Fig. 6a, second line in Table 2). The mean age of DCVs in the resident state thus increases gradually from the most proximal to the most distal bouton. This is consistent with the experimental findings of [7], who investigated the age of DCVs by marking the DCVs with a photoconvertible construct that changes fluorescence color depending on its age. It takes DCVs longer to reach distal boutons because DCVs are likely to be captured several times along the way.

Based on results reported in Fig. 3B of [7] for terminals with type II boutons, the ratio of red/green fluorescence between distal and proximal boutons is approximately 1.4. From Fig. 1D of [22] the increase of red(orange) to green fluorescence ratio in the first 10 h is approximately linear. Using the slope of the curve in Fig. 1D and given that in the first 5 h the red(orange)/green ratio increased approximately 4.2 times, we concluded that the difference between the age of DCVs in distal boutons and the age of DCVs in proximal boutons is approximately 2 h. This estimate should be taken with caution due to such factors as the difference in temperatures between the mammalian cells used in [22] and the fly cells used in [7] as well as the pH difference between mammalian and fly vesicles. These factors can influence the time course of aging of the timer protein. Also, fluorescence ratio measurements would be sensitive to the exact optics used and pH [23].

That said, the obtained estimate is consistent with 1.91 hours (15.46 hours in bouton 1 - 13.55 hours in bouton 26) as predicted by our model. It should also be noted that the animals used in the experiments described in [7] were 4-5 days old at the time of the experiments and that DCVs are not made during the first day [24]. Since Fig. 5a suggests that the oldest DCVs in boutons are approximately 60 hours old, DCV ages predicted by the model are within the ages of animals used in [7].

The mean age of anterograde transiting DCVs changes from 0.43 hours in the most proximal bouton (#26) to 6.81 hours in the most distal bouton (#1) (Fig. 6b, second line in Table 2). The increase is due to release of older, previously captured DCVs into the transiting state from the resident state. The mean age of retrograde transiting DCVs changes from 6.88 hours in the second most distal bouton (#2) to 9.37 hours in the most proximal bouton (#26) (Fig. 6c, second line in Table 2). The increase of the age of DCVs as the DCVs, after turning around in bouton 1, move retrogradely toward proximal boutons, is again due to release of older DCVs from the resident state.

An increase of parameter $\varepsilon$, which indicates the portion of DCVs that join the anterograde component after being re-released from the resident state, increases the mean DCV age in the resident states in boutons (compare first three rows in Table 2). This is because the release of older, previously captured DCVs into the anterograde transiting state increases the average DCV age in that state, and as DCVs continue moving toward more distal boutons, they can be recaptured, which in turn increases the age of DCVs in the resident state.

A decrease in the initial concentration of transiting DCVs, $n_{0,t}$, causes a reduction in the mean age of the resident DCVs in boutons (compare rows 2 and 4 in Table 2). This happens because in this situation, before reaching a particular bouton $M$, fewer DCVs have been captured previously and resided in more proximal boutons $(26,\ldots,M+1)$. This leads to younger DCVs in bouton $M$.

Table 2. The mean age of resident and transiting DCVs (in hours) in four representative boutons (#1, 5, 13, and 26) at steady-state. $\delta = 1$.

| $n_{0,t}$ | $\varepsilon$ | Mean age of resident DCVs in bouton | | | | Mean age of anterograde transiting DCVs in bouton | | | |
|---|---|---|---|---|---|---|---|---|---|
| | | #1 (h) | #5 (h) | #13 (h) | #26 (h) | #1,t (h) | #5,ta (h) | #13,ta (h) | #26,ta (h) |
| 1 | 0 | 11.32 | 11.45 | 12.32 | 13.38 | 2.67 | 2.18 | 1.30 | 0.08 |
| 1 | 0.8 | 15.46 | 15.47 | 14.98 | 13.55 | 6.81 | 6.51 | 4.69 | 0.43 |
| 1 | 1 | 16.24 | 16.23 | 15.53 | 13.59 | 7.59 | 7.32 | 5.37 | 0.51 |
| 0.1 | 0.8 | 13.21 | 13.22 | 12.77 | 11.58 | 4.56 | 4.52 | 3.29 | 0.30 |

| $n_{0,t}$ | $\varepsilon$ | Mean age of retrograde transiting DCVs in bouton | | | |
|---|---|---|---|---|---|
| | | #2,tr (h) | #5,tr (h) | #13,tr (h) | #26,tr (h) |
| 1 | 0 | 2.79 | 3.40 | 6.04 | 9.37 |
| 1 | 0.8 | 6.88 | 7.12 | 7.96 | 9.37 |
| 1 | 1 | 7.66 | 7.85 | 8.38 | 9.37 |
| 0.1 | 0.8 | 4.57 | 4.62 | 4.95 | 5.54 |

The sensitivity analysis of the mean age of DCVs in boutons to $n_{0,t}$ and $\varepsilon$ is presented in section S5.

## 4. Discussion, limitations of the model and future directions

Our model predicts that in type II terminals proximal boutons are filled before distal boutons (Fig. 3b). This prediction agrees with [7]. Resident DCVs exhibit a bimodal age density distribution (Fig. 5a), which is explained by the fact that DCVs are captured into the resident state from two pools: the anterograde transiting pool that contains younger vesicles and the retrograde transiting pool that contains older vesicles (Fig. 2). If it is assumed that captured DCVs, after spending some time is the resident state, are re-released back to circulation ($\delta = 1$), our modeling results also show that resident DCVs are older in distal boutons than in proximal boutons (Fig. 6). The mean age of DCVs in the resident state of the most distal bouton is approximately 15.5 hours while in the most proximal bouton it is approximately 13.5 hours (Table 2, line 2). The difference between the mean age of DCVs in proximal (younger DCVs) and distal (older DCVs) boutons (Fig. 6a) is explained by the fact that as they travel to distal boutons, for $\delta = 1$ DCVs experience several capture-re-release cycles along the way, spending some time in the resident state before being re-released to the transiting state.

The situation is different if DCVs are destroyed in boutons or leave the boutons by exocytosis ($\delta = 0$). In this case, at steady-state, resident DCVs in more proximal boutons are older than in more distal boutons (Table S4). Since this contradicts experimental results reported in [7], the model prediction supports the scenario when most captured DCVs are released back to circulation instead of being destroyed in boutons.

The predicted prevalence of older organelles in distal boutons for $\delta = 1$ may give important clues for establishing molecular mechanisms of the degeneration of axons of dopaminergic neurons in PD. This may explain "dying back" degeneration of axons that begins in the distal axon [25,26]. The prevalence of older organelles in distal boutons is a consequence of large and extensive arbors of dopaminergic neurons [27]. Older organelles in such neurons may be subject to damage by reactive oxygen species [25].

The model generates predictions for the mean age of DCVs and the DCV age distribution, which could be tested experimentally, giving the model predictive value. The model thus suggests future experiments that could be used to validate the model's predictions.

Future development of the model should address the following. (i) Possible change in the direction of DCV transport after DCVs have exited the terminal should be accounted for [5]. This would shift the age distribution in the terminal toward older DCVs. (ii) The alteration of capture efficiency by activity should also be included in the model [5,6,28]. (iii) Ideally, the delay of 300s in bouton 1 required to change anterograde to retrograde motors should be applied to each vesicle arriving to the bouton 1. The utilization of $H\left[t - t_1\right]$ function in Eq. (11)-(13) only provides an initial delay of the retrograde flux by 300s. This factor will not affect any DCV arriving after $t_1$ time. Future models should overcome this limitation of the compartmental model, which can be done by treating DCVs as individual vesicles [29]. The utilization of a discrete model would also provide much more detailed information on the vesicles trajectories and age. (iv) Recent experimental results indicate that DCVs release some of their contents by kiss and run exocytosis, in contrast to traditional full collapse exocytosis, which fully empties DCVs [30]. This should also be addressed in future model development.

The current version of the model assumes that at $t = 0$, DCVs start flowing into an empty type II terminal, which is assumed to be post-development. Experiments reported in [7] investigated animals at the end of their third instar larva stage, before the animals retracted their neurons to start the transformation to adulthood. In order to better simulate the experimental results, future versions of the model should consider coupling DCV transport with terminal development as the animals grow to reach the third instar larva stage. Accounting for the terminal growth could affect the predicted mean age of DCVs in distal boutons. This is because older boutons will already be filled with DCVs, which will reduce the chances for DCVs to get captured (and subsequently re-released) as they travel toward distal boutons. In particular, the model should address whether type II boutons are growing by adding new boutons either at the ends of axons or in-between existing boutons. This may be important because DCVs may bypass fully occupied (older) boutons, but be intensively captured in new boutons, which would affect the order in which the boutons are filled. The discussed questions would be interesting to address through a combination of modeling and experimentation in future research.

**Data accessibility.** Additional data accompanying this paper are available in the Supplemental Materials.

**Authors' contributions.** IAK and AVK contributed equally to the performing of computational work and article preparation.

**Competing interests.** We have no competing interests.

**Funding statement.** AVK acknowledges the support of the Alexander von Humboldt Foundation through the Humboldt Research Award and funding from the National Science Foundation (award CBET-1642262). IAK acknowledges the fellowship support of the Paul and Daisy Soros Fellowship for New Americans and the NIH/National Institutes of Mental Health (NIMH) Ruth L. Kirchstein NRSA (F30 MH122076-01).

**Acknowledgments.** We are grateful to Holger Metzler for careful reading of our manuscript and for his valuable comments and suggestions concerning the implementation of the method for computing the mean age of DCVs.

# References

1. van den Pol AN. 2012 Neuropeptide transmission in brain circuits. *Neuron* **76,** 98-115. (doi:10.1016/j.neuron.2012.09.014)

2. Li C, Wu X, Liu S, Zhao Y, Zhu J, Liu K. 2019 Roles of neuropeptide Y in neurodegenerative and neuroimmune diseases. *Frontiers in Neuroscience* **13,** 869. (doi:10.3389/fnins.2019.00869)

3. Lim A, Rechtsteiner A, Saxton WM. 2017 Two kinesins drive anterograde neuropeptide transport. *Mol. Biol. Cell* **28,** 3542-3553. (doi:10.1091/mbc.E16-12-0820)

4. Shakiryanova D, Tully A, Levitan ES. 2006 Activity-dependent synaptic capture of transiting peptidergic vesicles. *Nat. Neurosci.* **9,** 896-900. (doi:10.1038/nn1719)

5. Wong MY, Zhou C, Shakiryanova D, Lloyd TE, Deitcher DL, Levitan ES. 2012 Neuropeptide delivery to synapses by long-range vesicle circulation and sporadic capture. *Cell* **148,** 1029-1038. (doi:10.1016/j.cell.2011.12.036)

6. Bulgari D, Zhou C, Hewes RS, Deitcher DL, Levitan ES. 2014 Vesicle capture, not delivery, scales up neuropeptide storage in neuroendocrine terminals. *Proc. Natl. Acad. Sci. U. S. A.* **111,** 3597-3601. (doi:10.1073/pnas.1322170111)

7. Tao J, Bulgari D, Deitcher DL, Levitan ES. 2017 Limited distal organelles and synaptic function in extensive monoaminergic innervation. *J. Cell. Sci.* **130,** 2520-2529. (doi:10.1242/jcs.201111)

8. Bolam JP, Pissadaki EK. 2012 Living on the edge with too many mouths to feed: Why dopamine neurons die. *Movement Disorders* **27,** 1478-1483. (doi:10.1002/mds.25135)

9. Saxton WM, Hollenbeck PJ. 2012 The axonal transport of mitochondria. *J. Cell. Sci.* **125,** 2095-2104. (doi:10.1242/jcs.053850)

10. Kuznetsov IA, Kuznetsov AV. 2013 A compartmental model of neuropeptide circulation and capture between the axon soma and nerve terminals. *International Journal for Numerical Methods in Biomedical Engineering* **29,** 574-585. (doi:10.1002/cnm.2542)

11. Kuznetsov IA, Kuznetsov AV. 2015 Can an increase in neuropeptide production in the soma lead to DCV circulation in axon terminals with type III en passant boutons? *Math. Biosci.* **267,** 61-78.

12. Kuznetsov IA, Kuznetsov AV. 2015 Modeling neuropeptide transport in various types of nerve terminals containing en passant boutons. *Mathematical Biosciences* **261,** 27-36.

13. Kuznetsov IA, Kuznetsov AV. 2018 Simulating reversibility of dense core vesicles capture in en passant boutons: Using mathematical modeling to understand the fate of dense core vesicles in en passant boutons. *Journal of Biomechanical Engineering-Transactions of the ASME* **140,** 051004. (doi:10.1115/1.4038201)

14. Kuznetsov IA, Kuznetsov AV. 2019 Modelling transport and mean age of dense core vesicles in large axonal arbours. *Proceedings of the Royal Society A* **475,** 20190284. (doi:10.1098/rspa.2019.0284)

15. Anderson DH. 1983 *Compartmental Modeling and Tracer Kinetics.* Berlin: Springer.

16. Jacquez JA. 1985 *Compartmental Analysis in Biology and Medicine.* 2nd ed. ed., University of Michigan Press, Ann Arbor, MI.

17. Jacquez JA, Simon CP. 1993 Qualitative theory of compartmental-systems. *SIAM Rev* **35,** 43-79. (doi:10.1137/1035003)

18. Metzler H, Muller M, Sierra CA. 2018 Transit-time and age distributions for nonlinear time-dependent compartmental systems. *Proc. Natl. Acad. Sci. U. S. A.* **115,** 1150-1155. (doi:10.1073/pnas.1705296115)

19. Metzler H, Sierra CA. 2018 Linear autonomous compartmental models as continuous-time markov chains: Transit-time and age distributions. *Mathematical Geosciences* **50,** 1-34. (doi:10.1007/s11004-017-9690-1)

20. Rasmussen M, Hastings A, Smith MJ, Agusto FB, Chen-Charpentier BM, Hoffman FM, Jiang J, Todd-Brown KEO, Wang Y, Wang Y, Luo Y. 2016 Transit times and mean ages for nonautonomous and autonomous compartmental systems. *J. Math. Biol.* **73,** 1379-1398. (doi:10.1007/s00285-016-0990-8)

21. Levitan ES. 2017 Personal communication.

22. Tsuboi T, Kitaguchi T, Karasawa S, Fukuda M, Miyawaki A. 2010 Age-dependent preferential dense-core vesicle exocytosis in neuroendocrine cells revealed by newly developed monomeric fluorescent timer protein. *Mol. Biol. Cell* **21,** 87-94. (doi:10.1091/mbc.E09-08-0722)

23. Levitan ES. 2020 Personal communication.

24. Levitan ES. 2019 Personal communication.

25. Burke RE, O'Malley K. 2013 Axon degeneration in Parkinson's disease. *Exp. Neurol.* **246,** 72-83. (doi:10.1016/j.expneurol.2012.01.011)

26. Tagliaferro P, Burke RE. 2016 Retrograde axonal degeneration in Parkinson disease. *Journal of Parkinsons Disease* **6,** 1-15. (doi:10.3233/JPD-150769)

27. Mamelak M. 2018 Parkinson's disease, the dopaminergic neuron and gammahydroxybutyrate. *Neurology and Therapy* **7,** 5-11. (doi:10.1007/s40120-018-0091-2)

28. Cavolo SL, Bulgari D, Deitcher DL, Levitan ES. 2016 Activity induces Fmr1-sensitive synaptic capture of anterograde circulating neuropeptide vesicles. *Journal of Neuroscience* **36,** 11781-11787. (doi:10.1523/JNEUROSCI.2212-16.2016)

29. Lai X, Brown A, Xue C. 2018 A stochastic model that explains axonal organelle pileups induced by a reduction of molecular motors. *Journal of the Royal Society Interface* **15,** 20180430. (doi:10.1098/rsif.2018.0430)

30. Wong MY, Cavolo SL, Levitan ES. 2015 Synaptic neuropeptide release by dynamin-dependent partial release from circulating vesicles. *Mol. Biol. Cell* **26,** 2466-2474. (doi:10.1091/mbc.E15-01-0002)

31. Levitan ES. 2018 Personal communication.

32. Beck JV, Arnold KJ. 1977 *Parameter Estimation in Science and Engineering.* New York: Wiley.

33. Zadeh KS, Montas HJ. 2010 A class of exact solutions for biomacromolecule diffusion-reaction in live cells. *J. Theor. Biol.* **264,** 914-933. (doi:10.1016/j.jtbi.2010.03.028)

34. Zi Z. 2011 Sensitivity analysis approaches applied to systems biology models. *IET Systems Biology* **5,** 336-346. (doi:10.1049/iet-syb.2011.0015)

35. Kuznetsov IA, Kuznetsov AV. 2019 Investigating sensitivity coefficients characterizing the response of a model of tau protein transport in an axon to model parameters. *Comput. Methods Biomech. Biomed. Engin.* **22,** 71-83. (doi:10.1080/10255842.2018.1534233)

36. Kacser H, Burns J, Fell D. 1995 The control of flux. *Biochem. Soc. Trans.* **23,** 341-366. (doi:10.1042/bst0230341)

**Figure captions**

Fig. 1. (a) A schematic diagram of a neuron with an axon whose arbor splits into four branches. Each branch is assumed to contain 26 boutons. Boutons are numbered starting with the most distal (#1) to the most proximal (#26), consistent with the convention used in [5]. (b) A magnified portion of the terminal containing boutons 20, 19, 18, and 17. Transiting and resident DCVs as well as DCV fluxes between the boutons are displayed. Block arrows show capture of transiting DCVs into the resident state. The rates of capture are also indicated.

Fig. 2. A diagram of a compartmental model showing transport in the transiting and resident states in the terminal. Arrows show DCV exchange between the transiting states in adjacent boutons, DCV capture into the resident state and re-release from the resident state, as well as DCV destruction in the resident state. DCVs entering the terminal are assumed to have zero age.

Fig. 3. (a) Saturated DCV concentrations in the resident state in boutons, from the most proximal (#26) to the most distal (#1) bouton. Estimated values of saturated concentrations, calculated using Eqs. (S4) and (S5), are compared with numerically obtained values of these concentrations (obtained from the numerical solution of Eqs. (1)-(8) with boundary conditions (14), (15) at $t \to \infty$). $n_{0,t} = 1$ and $\varepsilon = 0.8$. Results are independent of the value of $\delta$. (b) DCV concentrations in the resident state at different times. $\delta = 1$, $n_{0,t} = 1$, and $\varepsilon = 0.8$.

Fig. 4. The buildup toward steady-state: concentrations of DCVs in the resident state in various boutons. (a) Boutons 26 through 14. (b) Boutons 13 through 1. $\delta = 1$, $n_{0,t} = 1$, and $\varepsilon = 0.8$.

Fig. 5. (a) Age density of DCVs in the resident state in various boutons at steady-state. (b) Age density of DCVs in the anterograde transiting state in various boutons at steady-state. (c) Age density of DCVs in the retrograde transiting state in various boutons at steady-state. $\delta = 1$, $n_{0,t} = 1$, and $\varepsilon = 0.8$. Age density is shown in every second bouton to make the figures less cluttered.

Fig. 6. (a) Mean age of resident DCVs in various boutons versus time. (b) Mean age of anterograde transiting DCVs in various boutons versus time. (c) Mean age of retrograde transiting DCVs in various boutons versus time. $\delta = 1$, $n_{0,t} = 1$, and $\varepsilon = 0.8$.

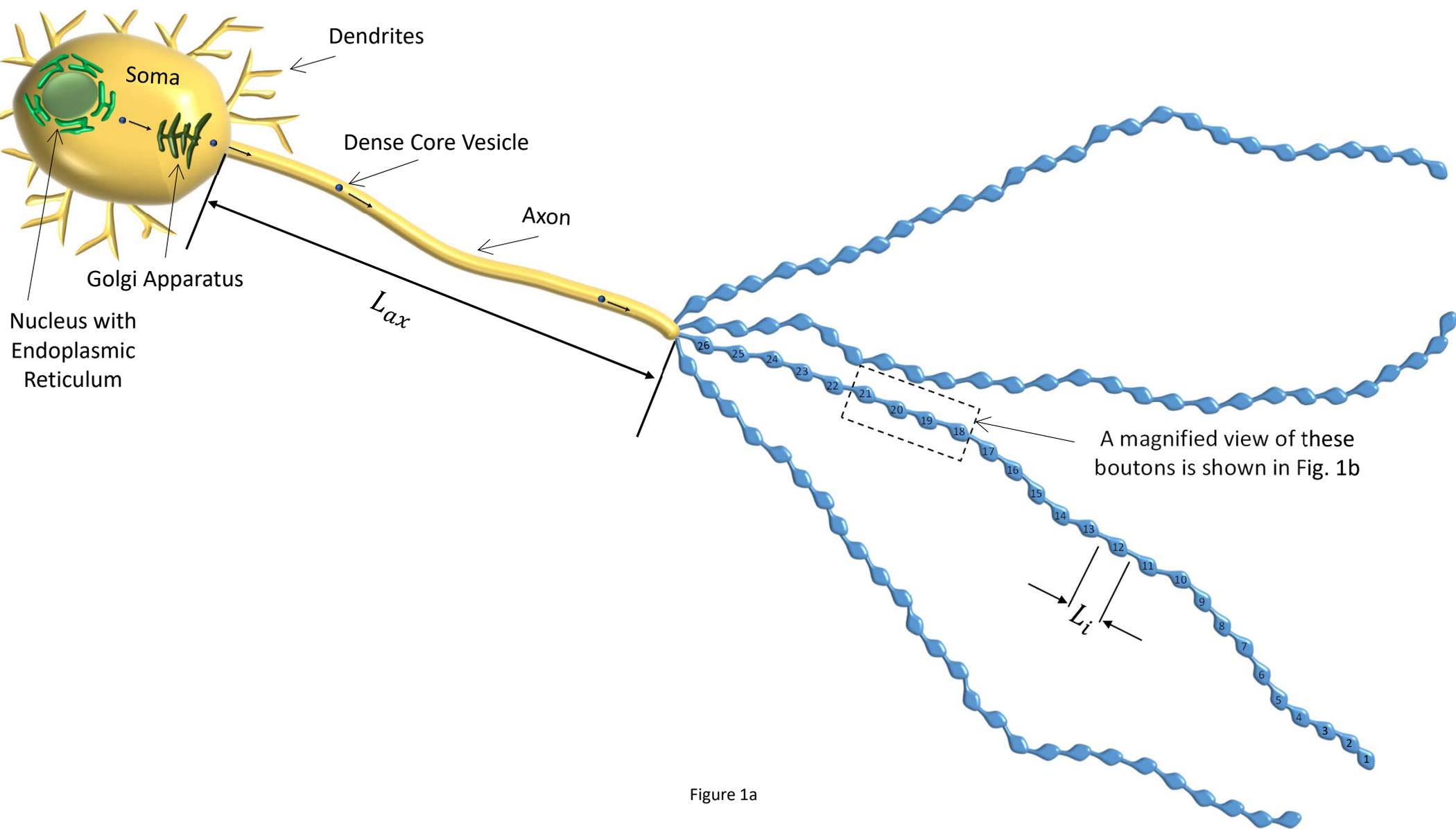

A magnified view of these boutons is shown in Fig. 1b

Figure 1a

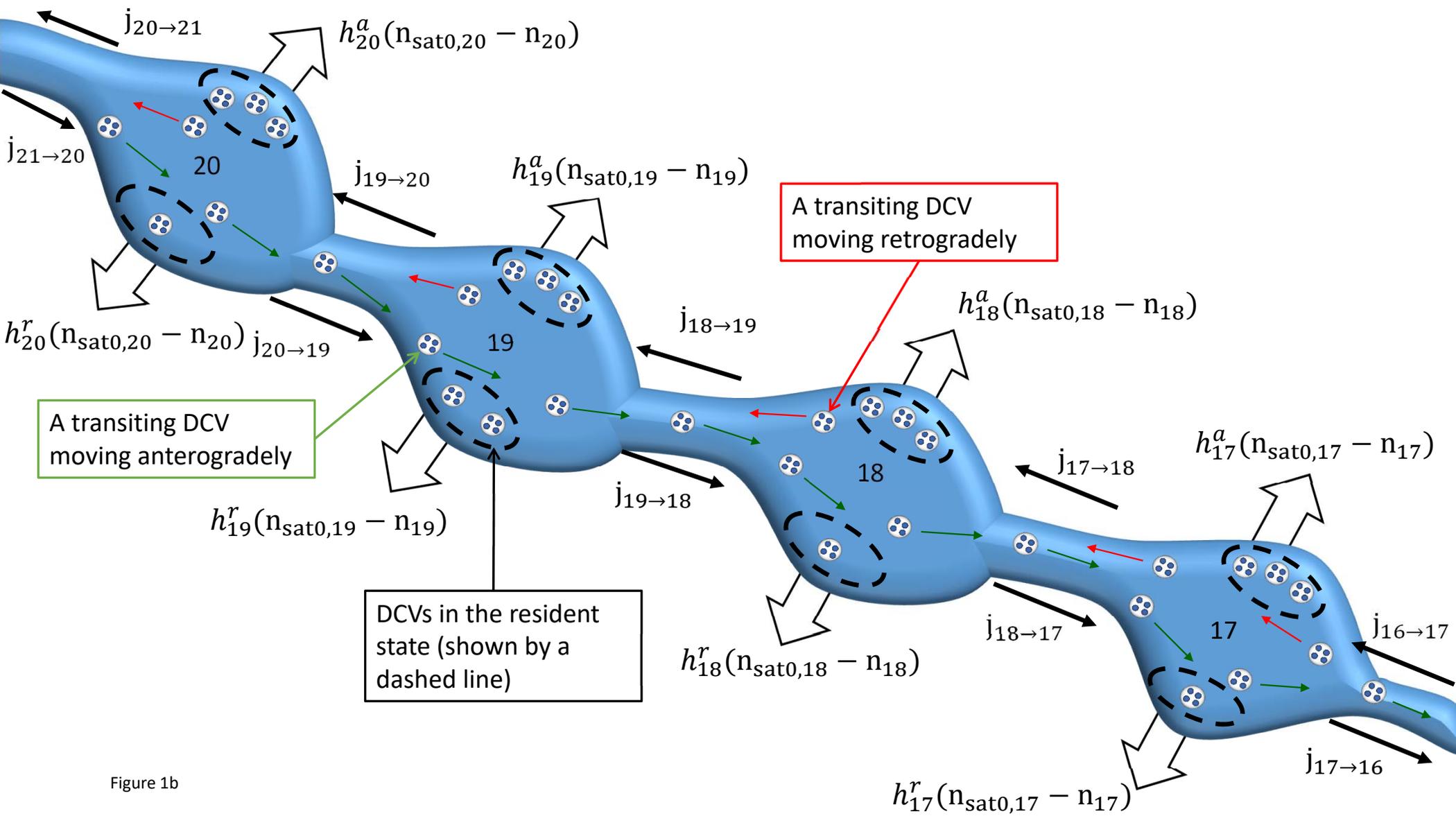

$j_{20 \to 21}$

$h_{20}^a(n_{sat0,20} - n_{20})$

$j_{21 \to 20}$

$j_{19 \to 20}$

$h_{19}^a(n_{sat0,19} - n_{19})$

A transiting DCV moving retrogradely

$h_{20}^r(n_{sat0,20} - n_{20})$ $j_{20 \to 19}$

A transiting DCV moving anterogradely

$h_{18}^a(n_{sat0,18} - n_{18})$

$j_{18 \to 19}$

$h_{19}^r(n_{sat0,19} - n_{19})$

$j_{17 \to 18}$

$h_{17}^a(n_{sat0,17} - n_{17})$

$j_{19 \to 18}$

DCVs in the resident state (shown by a dashed line)

$h_{18}^r(n_{sat0,18} - n_{18})$

$j_{18 \to 17}$

$j_{16 \to 17}$

$h_{17}^r(n_{sat0,17} - n_{17})$

$j_{17 \to 16}$

Figure 1b

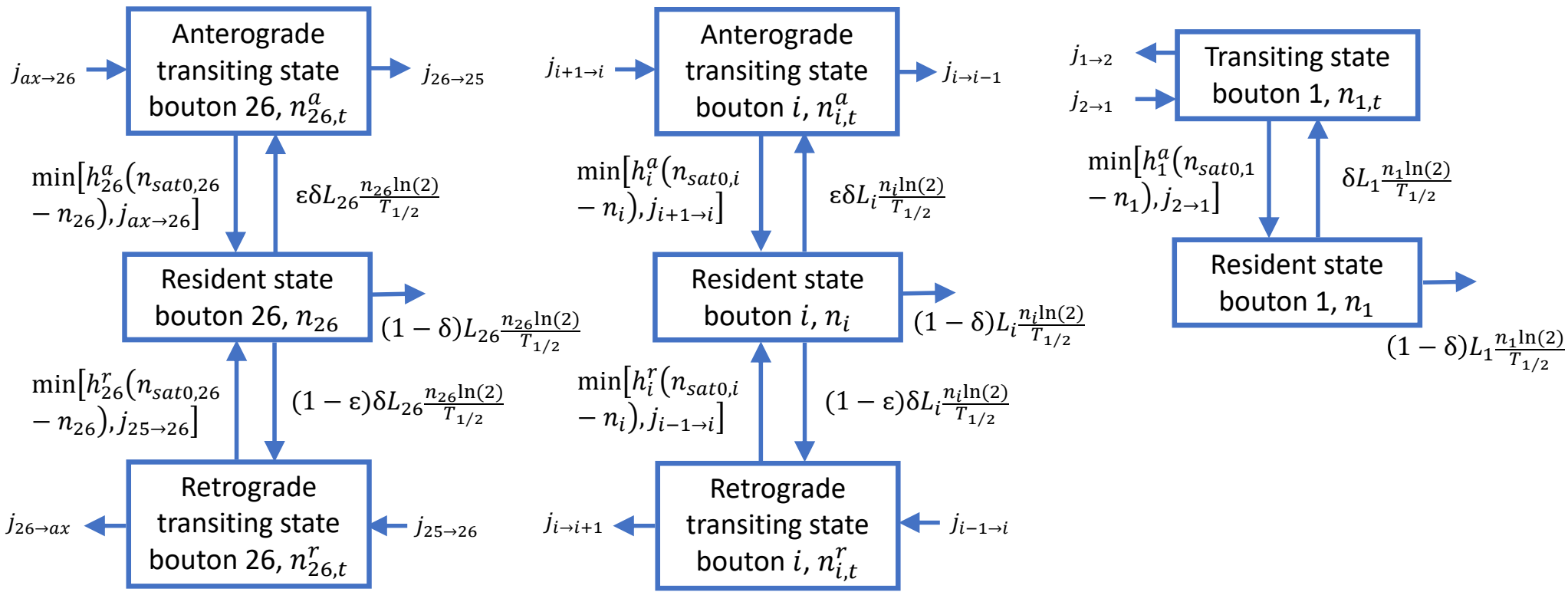

Figure 2

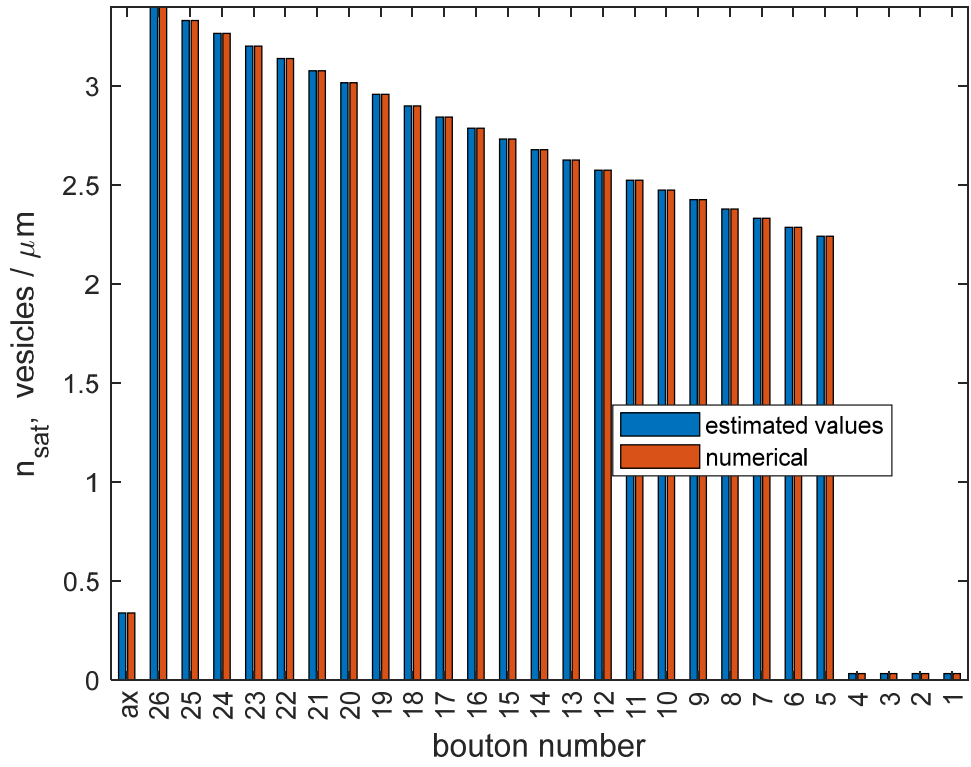

(a)

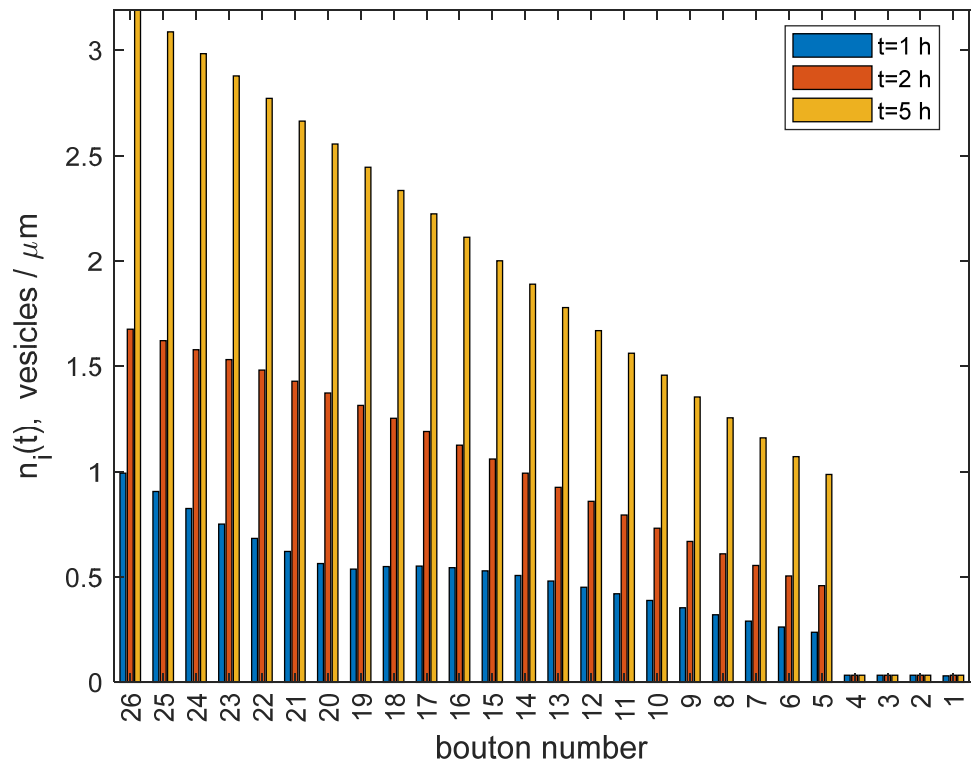

(b)

Figure 3

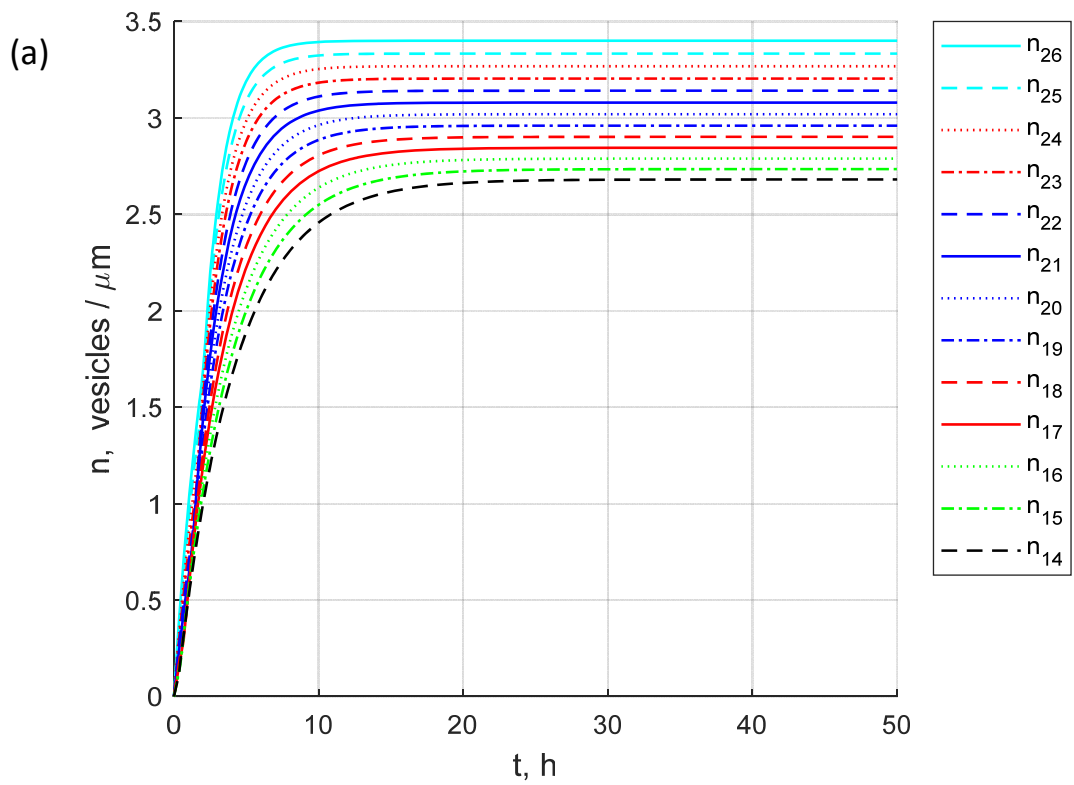

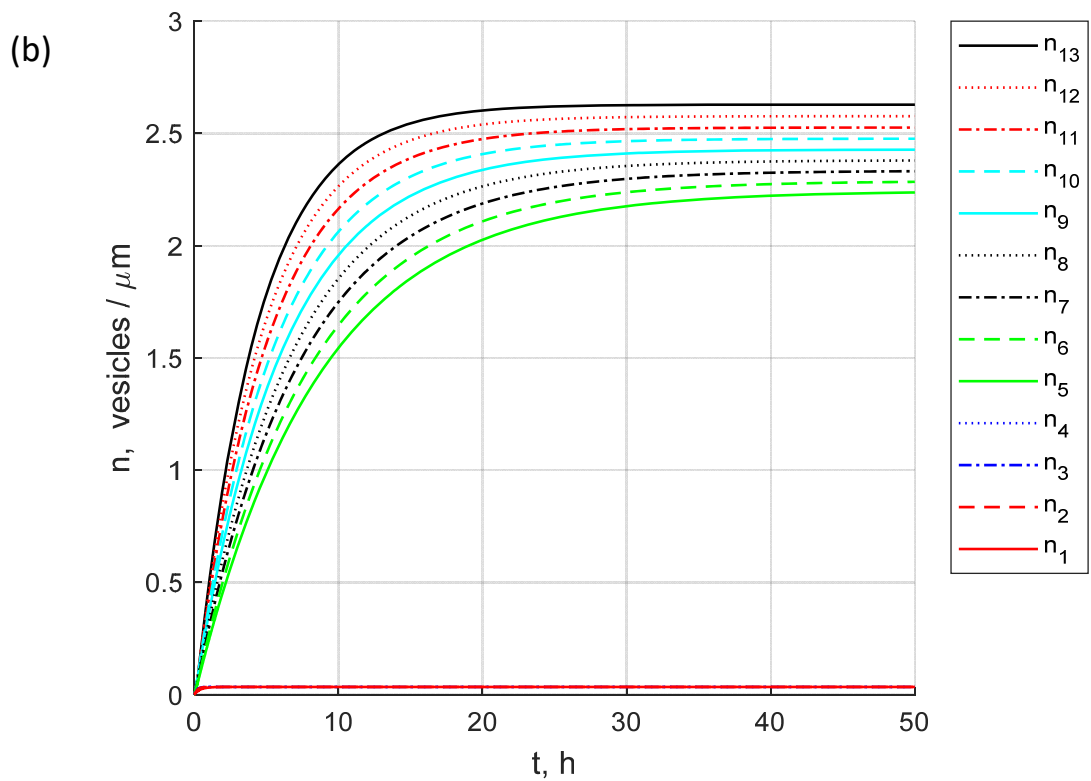

Figure 4

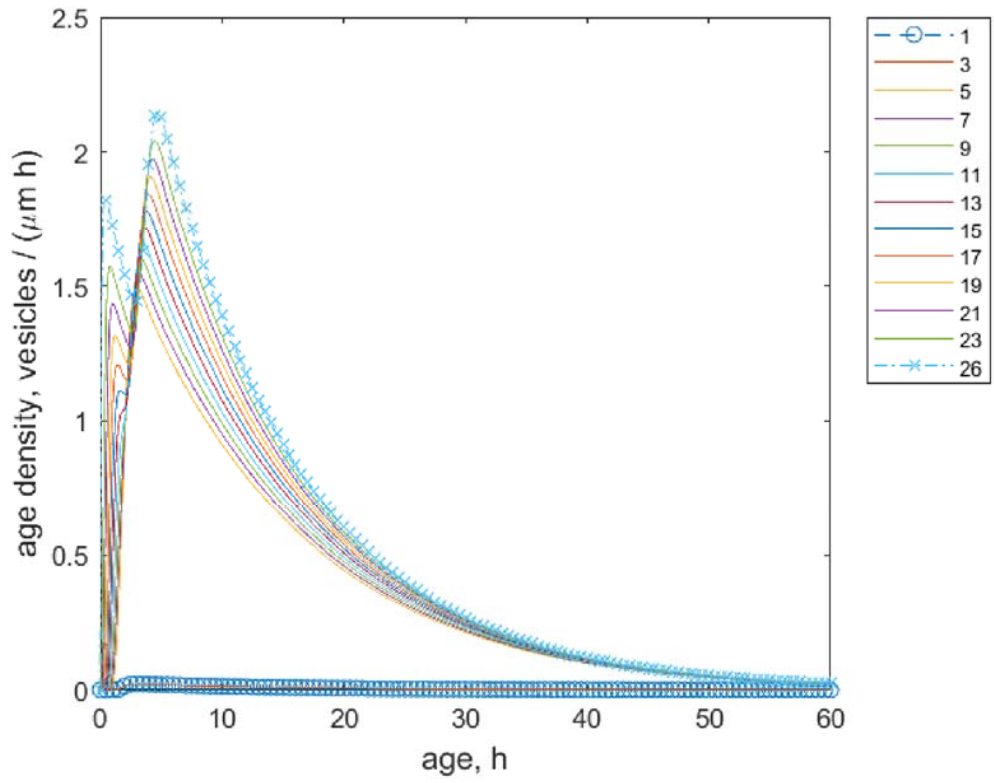

(a)

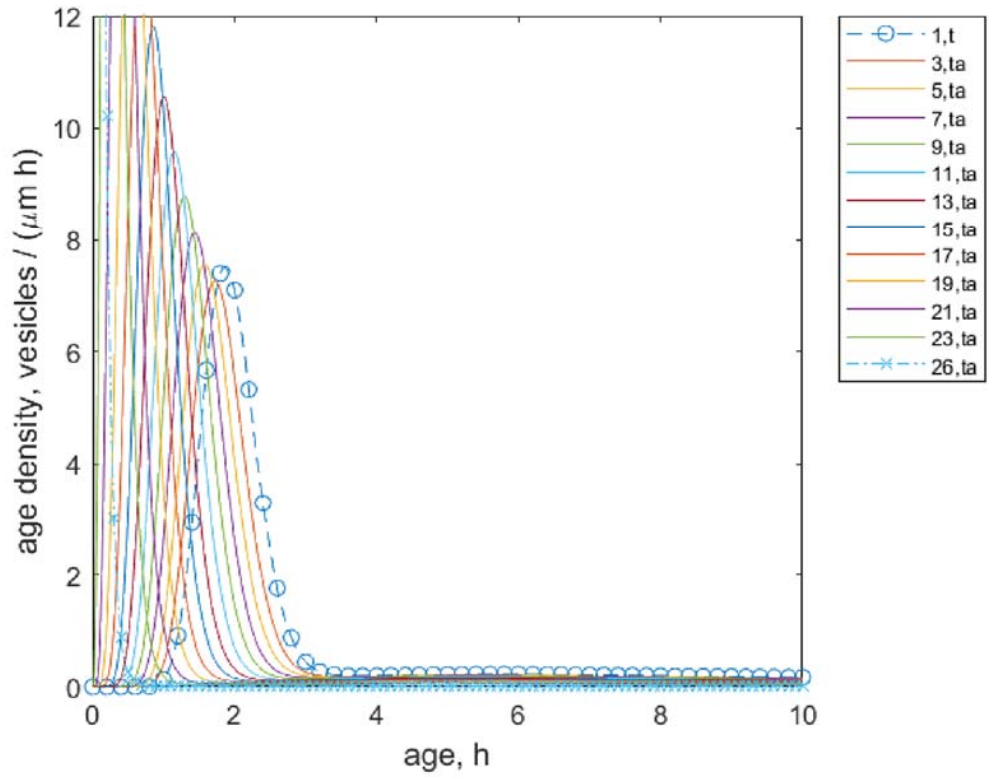

(b)

Figure 5a,b

(c)

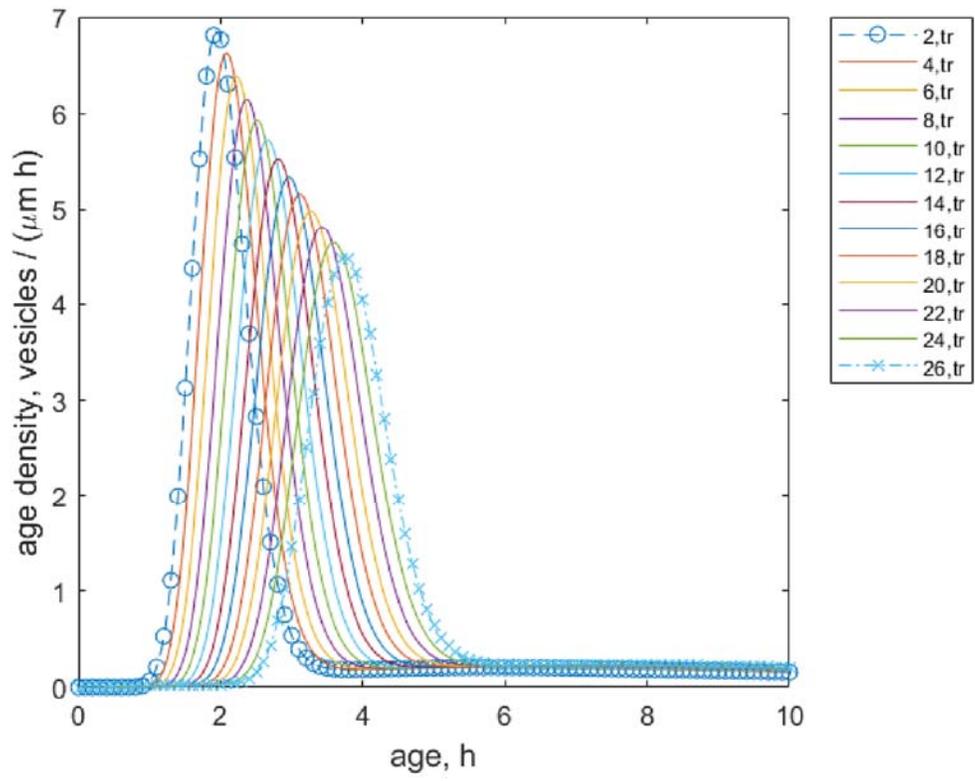

Figure 5c

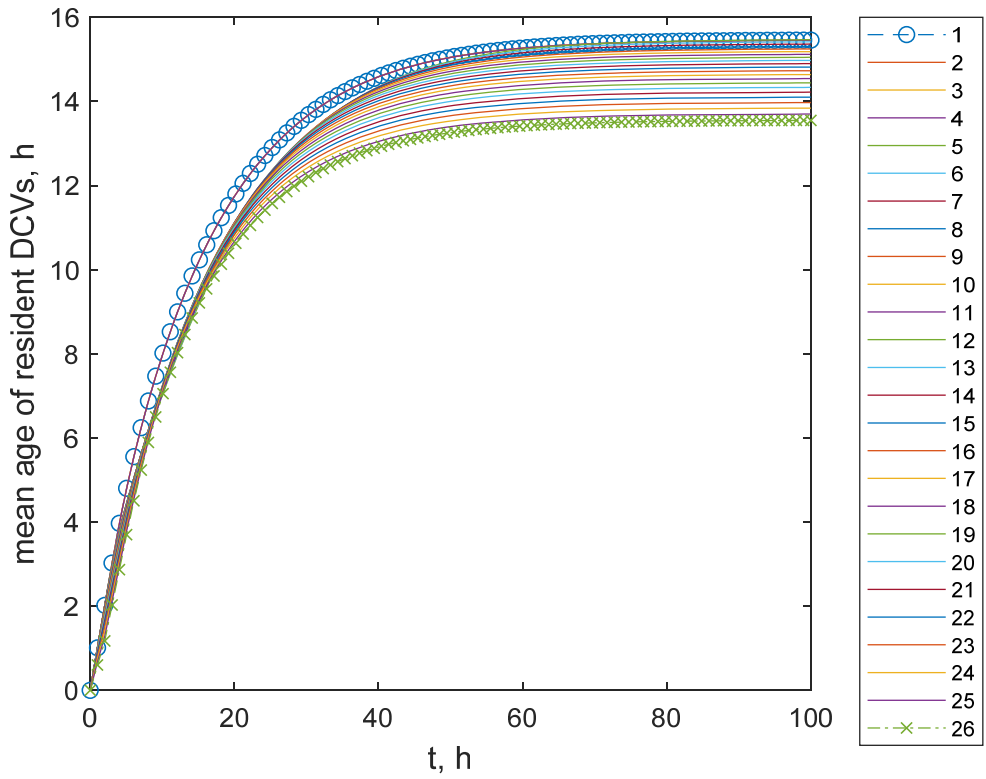

(a)

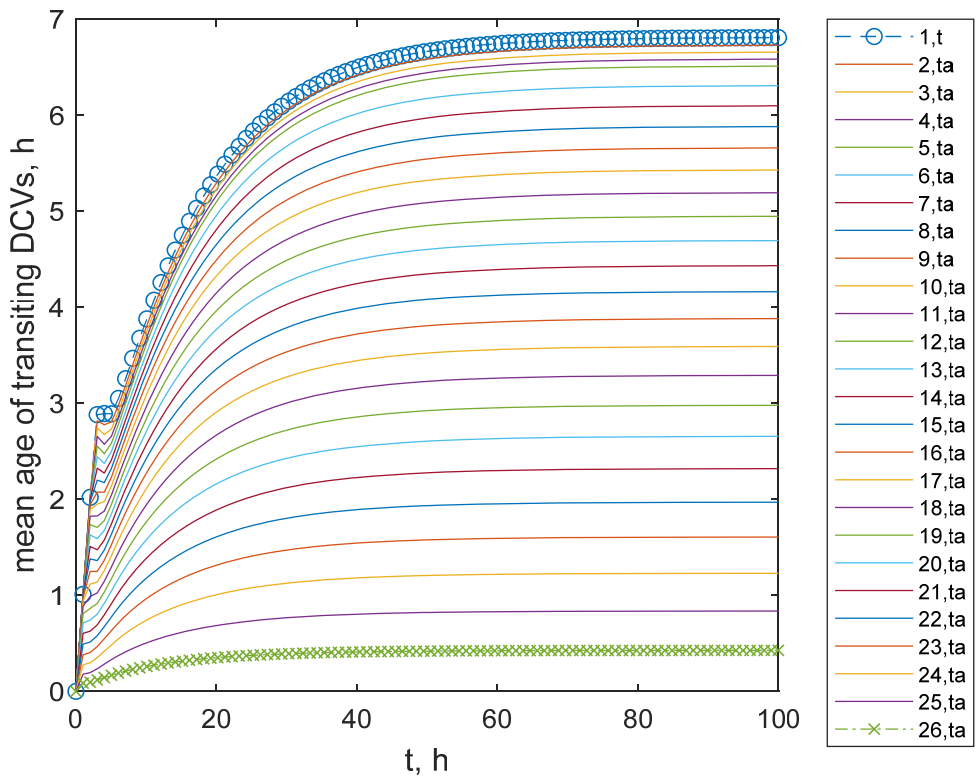

(b)

Figure 6a,b

(c)

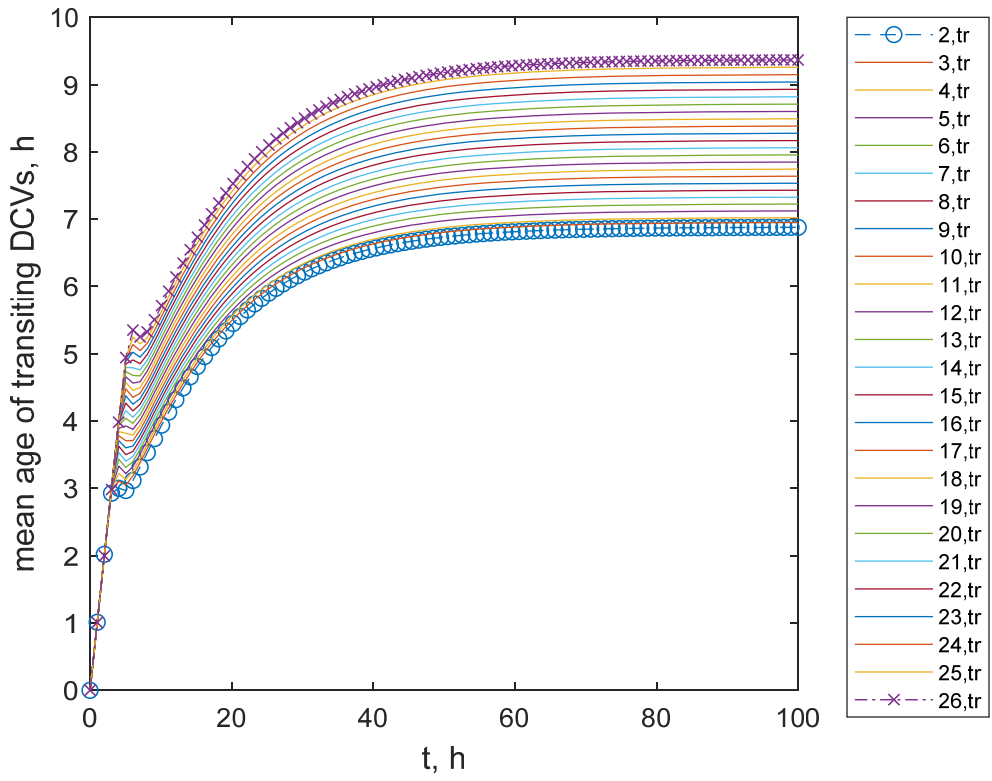

Figure 6c

**How old are dense core vesicles residing in en passant boutons: Simulation of the mean age of dense core vesicles in axonal arbors accounting for resident and transiting vesicle populations**

I. A. Kuznetsov[(a), (b)] and A. V. Kuznetsov[(c)]

[(a)]Perelman School of Medicine, University of Pennsylvania, Philadelphia, PA 19104, USA

[(b)]Department of Bioengineering, University of Pennsylvania, Philadelphia, PA 19104, USA

[(c)]Dept. of Mechanical and Aerospace Engineering, North Carolina State University,

Raleigh, NC 27695-7910, USA; e-mail: avkuznet@ncsu.edu

## Supplementary Material

### S1. Modeling the concentration of DCVs in the axon

Variation of the DCV concentration in the axon can be simulated as suggested in [14]. The only DCVs present in the axon are transiting DCVs. Assuming that the terminal splits into four identical branches (Fig. 1a), the conservation of transiting DCVs in the axon can be stated as follows:

$$L_{ax} \frac{dn_{ax}}{dt} = j_{soma \to ax} + 4 j_{26 \to ax} - 4 j_{ax \to 26} - L_{ax} \frac{n_{ax} \ln(2)}{T_{1/2,ax}}, \qquad (S1)$$

where $n_{ax}$ is the average concentration of DCVs in the axon, $L_{ax}$ is the length of the axon (Fig. 1a), $j_{soma \to ax}$ is the net DCV flux from the soma into the axon (the DCV production rate in the soma minus the DCV destruction rate in the somatic lysosomes), and $T_{1/2,ax}$ is the half-life of DCVs in the axon. In Eq. (S1) we assumed that DCV half-lives are likely different in the axon ($T_{1/2,ax}$) and boutons ($T_{1/2}$, see Eqs. (1)-(8)). One could imagine that the biochemical marking of vesicle proteins for degradation (e.g., by ubiquitination) could be more robust in boutons. Also, neuropeptides may leave the boutons by exocytosis. These phenomena would shorten the half-life of neuropeptides in boutons. On the other hand, DCVs are likely to be protected from degradation during their transport in the axon; otherwise, in long axons most DCVs would be destroyed before reaching the terminal.

The flux from the axon to the most proximal terminal is assumed to be proportional to the average DCV concentration in the axon:

$$j_{ax \to 26} = h_{ax} n_{ax},$$ (S2)

where $h_{ax}$ is the mass transfer coefficient that characterizes the rate at which DCVs leave the axon and enter the most proximal bouton.

For the purpose of defining the initial conditions for Eq. (S1), it can be assumed that initially the axon is filled to saturation:

$$n_{ax}(0) = n_{sat,ax}.$$ (S3)

## S2. Estimation of values of parameters involved in the model

### S2.1. Two groups of parameters involved in the model, divided by the method how they are estimated

We divided the model parameters into two groups. The parameters belonging to the first group ($j_{ax \to 26}$, $L_1$,..., $L_{26}$, $L_{ax}$, $t_1$, $T_{1/2}$, $T_{1/2,ax}$, and $\varpi$) were estimated using data found in published literature or assumed on physical grounds. These parameters are reported in Table S1. The parameters belonging to the second group ($n_{sat,ax}$, $n_{sat,1}$,..., $n_{sat,26}$, $h_1$,..., $h_{26}$, $h_{in}$, $n_{sat0,1}$,..., $n_{sat0,26}$, $n_{sat,ax}$, and $j_{soma}$) were found by stating the conservation of DCVs at $t = 0$ and at steady-state. These parameters are reported in Tables S2 and S3. Below, we explain how we estimated values of parameters belonging to the second group.

Table S1. Parameters of the model that we estimated based on values found in the literature or assumed on physical grounds.

| Symbol | Definition | Units | Estimated value(s) or range | Reference(s) |
|--------|------------|-------|-----------------------------|--------------|
| $a$ | Parameter characterizing the saturated DCV concentration in the resident state, see Eq. (S1) | | 1.02 | |
| $j_{ax \to 26}$ | Flux from the axon into the most proximal bouton | vesicles/ s | 0.0333 | [3,7] |
| $L_i$ | Length of a compartment occupied by bouton $i$ ($i$= 1,..., 26), defined in Fig. 1a | μm | 10 | Levitan ES. 2018, personal |

| | | | | communication |
|---|---|---|---|---|
| $n_{0,t}$ | DCV concentration in the transiting state at $t=0$ | vesicles/μm | 0.1, 1 | |
| $t_1$ | Average time that it takes DCVs to change the direction in the most distal bouton, if they are not captured (probably involves the change of molecular motors that propel DCVs) | s | 300 | [3], Levitan ES. 2017, personal communication |
| $T_{1/2}$ | Half-life or half-residence time of DCVs in the resident state in boutons | s | $2.16 \times 10^4$ | [5] |
| $\varepsilon$ | The model requires postulating how DCVs released from the resident state are split between the anterograde and retrograde components. We assumed that the portion of released DCVs that join the anterograde component is $\varepsilon$ while the portion of released DCVs that join the retrograde component is $(1-\varepsilon)$. | | 0-1 | |
| $\delta$ | The model requires postulating what portion of captured DCVs are destroyed in boutons and what portion are released back to circulation. We assumed that $\delta$ is the portion of DCVs released from the captured state back to the transiting state and $(1-\delta)$ is the portion of captured DCVs that are eventually destroyed in boutons. | | 0-1 | |
| $\lambda$ | Parameter characterizing the magnitude of the drop of the saturated DCV concentration in the resident state in the farthest boutons | | 0.01 | |
| $\varpi$ | DCV capture efficiency, defined as the percentage of DCVs captured in an empty bouton when DCVs pass the bouton. In all boutons, except the most distal bouton, capture occurs | | 0.1 | [7] |

| | twice, when DCVs pass the bouton in anterograde and in retrograde directions. | | | |
|---|---|---|---|---|

## S2.2. Estimation of saturated concentrations of DCVs in the resident state in boutons, $n_{sat,i}$ ($i = 1,…,26$)

We considered type II terminals, which were studied in [7]. These terminals have approximately 80 boutons per muscle, which are distributed along 3 or 4 branches (Levitan ES. 2018, personal communication). Based on this, we estimated that there are approximately 26 boutons per single branch. There are approximately 34 DCVs in the most proximal type II bouton in the saturated state [7]. Results of [7] also suggest that the drop-off in the DCV content appears suddenly at the farthest ends of the arbor. To model this situation, we assumed the following steady-state distribution of DCVs in various boutons:

$$n_{sat,i} = (34 / L_i) / a^{26-i} = 3.4 / a^{26-i} \text{ vesicles/µm } (i = 5,…,26).$$  (S4)

In Eq. (S4) we used $a = 1.02$ to model a slow decrease of the DCV concentration toward the end of the arbor. In the four farthest boutons, we assumed a sudden drop-off of the DCV concentration:

$$n_{sat,i} = \lambda (34 / L_i) = 0.034 \text{ vesicles/µm } (i = 1,…,4),$$  (S5)

where $\lambda$ is a parameter characterizing the magnitude of the drop (see Table S1).

Eqs. (S4) and (S5) postulate a set capacity of the resident state in boutons for DCVs (this can be viewed as a fixed number of parking spaces in each bouton). This limits DCV accumulation in boutons and allows boutons with excess supply of vesicles to fill to a set amount. The remaining vesicles pass the bouton and continue traveling distally.

## S2.3. Estimation of mass transfer coefficients characterizing DCV capture into the resident state in boutons, $h_1$, $h_2$,…,$h_{26}$; and saturated concentrations of DCVs in boutons at infinite DCV half-life or half-residence time, $n_{sat0,1}$, $n_{sat0,2}$,…, $n_{sat0,26}$

Mass transfer coefficients characterizing DCV capture as they pass the boutons in anterograde and retrograde directions were assumed to be equal:

$$h_i^a = h_i^r = h_i \qquad (i=2,\ldots,26).$$ (S6)

We used two types of statements to estimate the values of $h_i$ and $n_{sat0,i}$ ($i=1,\ldots,26$). We first stated the DCV capture rate at $t=0$ in various boutons:

$$\varpi\, j_{ax\to26} = h_{26}\left(n_{sat0,26} - 0\right),$$ (S7)

…

$$\varpi\, j_{ax\to26}\left(1-\varpi\right)^{26-i} = h_i\left(n_{sat0,i} - 0\right),$$ (S8)

…

$$\varpi\, j_{ax\to26}\left(1-\varpi\right)^{25} = h_1\left(n_{sat0,1} - 0\right),$$ (S9)

where $\varpi$ is the capture efficiency, a parameter that characterizes the percentage of captured DCVs as they pass an empty bouton [7,14].

The second type of statements concerns the rate of DCV capture at steady-state. If half-life or half-residence time of resident DCVs is finite, the capture must still continue, even at steady-state:

$$2h_i\left(n_{sat0,i} - n_{sat,i}\right) = L_i \frac{n_{sat,i}\ln\left(2\right)}{T_{1/2}} \qquad (i=2,\ldots,26).$$ (S10)

DCVs can be captured when they pass anterogradely or retrogradely through the boutons, which explains the factor of two on the left-hand side of Eq. (S10). The only exception is bouton 1, which DCVs pass only once:

$$h_1\left(n_{sat0,1} - n_{sat,1}\right) = L_1 \frac{n_{sat,1}\ln\left(2\right)}{T_{1/2}}.$$ (S11)

We used Matlab's (Matlab R2019a, MathWorks, Natick, MA, USA) solver Solve, and we summarized the results in Tables S2 and S3.

Table S2. Mass transfer coefficients characterizing the rates of capture of DCVs into the resident state. The units of all $h_i$ in Table S2 are μm/s.

| $h_1$ | $h_2$ | $h_3$ | $h_4$ | $h_5$ | $h_6$ | $h_7$ |
|---|---|---|---|---|---|---|
| $6.71 \times 10^{-3}$ | $7.65 \times 10^{-3}$ | $8.52 \times 10^{-3}$ | $9.48 \times 10^{-3}$ | $1.98 \times 10^{-6}$ | $1.65 \times 10^{-5}$ | $3.23 \times 10^{-5}$ |

| $h_8$ | $h_9$ | $h_{10}$ | $h_{11}$ | $h_{12}$ | $h_{13}$ | $h_{14}$ |
|---|---|---|---|---|---|---|
| $4.95 \times 10^{-5}$ | $6.83 \times 10^{-5}$ | $8.87 \times 10^{-5}$ | $1.11 \times 10^{-4}$ | $1.35 \times 10^{-4}$ | $1.62 \times 10^{-4}$ | $1.90 \times 10^{-4}$ |

| $h_{15}$ | $h_{16}$ | $h_{17}$ | $h_{18}$ | $h_{19}$ | $h_{20}$ | $h_{21}$ |
|---|---|---|---|---|---|---|
| $2.22 \times 10^{-4}$ | $2.56 \times 10^{-4}$ | $2.93 \times 10^{-4}$ | $3.34 \times 10^{-4}$ | $3.78 \times 10^{-4}$ | $4.26 \times 10^{-4}$ | $4.78 \times 10^{-4}$ |

| $h_{22}$ | $h_{23}$ | $h_{24}$ | $h_{25}$ | $h_{26}$ |
|---|---|---|---|---|
| $5.35 \times 10^{-4}$ | $5.97 \times 10^{-4}$ | $6.65 \times 10^{-4}$ | $7.39 \times 10^{-4}$ | $8.19 \times 10^{-4}$ |

Table S3. Saturated concentrations of DCVs in boutons at infinite DCV half-life or at infinite DCV residence time. The units of $n_{sat0,i}$ in Table S3 are vesicles/µm.

| $n_{sat0,1}$ | $n_{sat0,2}$ | $n_{sat0,3}$ | $n_{sat0,4}$ | $n_{sat0,5}$ | $n_{sat0,6}$ | $n_{sat0,7}$ |
|---|---|---|---|---|---|---|
| $3.56 \times 10^{-2}$ | $3.47 \times 10^{-2}$ | $3.46 \times 10^{-2}$ | $3.46 \times 10^{-2}$ | $1.84 \times 10^{2}$ | $2.46 \times 10^{1}$ | $1.39 \times 10^{1}$ |

| $n_{sat0,8}$ | $n_{sat0,9}$ | $n_{sat0,10}$ | $n_{sat0,11}$ | $n_{sat0,12}$ | $n_{sat0,13}$ | $n_{sat0,14}$ |
|---|---|---|---|---|---|---|
| $1.01 \times 10^{1}$ | 8.14 | 6.96 | 6.18 | 5.64 | 5.24 | 4.94 |

| $n_{sat0,15}$ | $n_{sat0,16}$ | $n_{sat0,17}$ | $n_{sat0,18}$ | $n_{sat0,19}$ | $n_{sat0,20}$ | $n_{sat0,21}$ |
|---|---|---|---|---|---|---|
| 4.71 | 4.54 | 4.40 | 4.30 | 4.22 | 4.16 | 4.11 |

| $n_{sat0,22}$ | $n_{sat0,23}$ | $n_{sat0,24}$ | $n_{sat0,25}$ | $n_{sat0,26}$ |
|---|---|---|---|---|
| 4.08 | 4.06 | 4.06 | 4.06 | 4.07 |

Table S4. The mean age of resident and transiting DCVs (in hours) in four representative boutons (#1, 5, 13, and 26) at steady-state. $\delta = 0$.

| $n_{0,t}$ | $\varepsilon$ | Mean age of resident DCVs in bouton | | | | Mean age of anterograde transiting DCVs in bouton | | | |
|---|---|---|---|---|---|---|---|---|---|
| | | #1 (h) | #5 (h) | #13 (h) | #26 (h) | #1,t (h) | #5,ta (h) | #13,ta (h) | #26,ta (h) |
| 1 | 0.8 | 11.32 | 11.32 | 11.39 | 13.84 | 2.66 | 2.18 | 1.30 | 0.08 |

| $n_{0,t}$ | $\varepsilon$ | Mean age of retrograde transiting DCVs in bouton | | | |
|---|---|---|---|---|---|
| | | #2,tr (h) | #5,tr (h) | #13,tr (h) | #26,tr (h) |
| 1 | 0.8 | 2.78 | 3.14 | 4.16 | 6.29 |

## S3. Numbering of the compartments

The first 26 compartments correspond to resident states in boutons, compartment 27 corresponds to the only transiting state in bouton 1, compartments 28,…,52 correspond to transiting-anterograde states in boutons $j - 26$, where $j$ is the compartment number, and compartments 53,…,77 correspond to transiting-retrograde states in boutons $j - 51$ (Fig. 2).

## S4. Numerical solution

The solution procedure involves solving initial value problems for three sets of ODEs. We solved all three sets of ODEs numerically using Matlab's solver, ODE45 (Matlab R2019a, MathWorks, Natick, MA, USA). We set the error tolerance parameters, RelTol and AbsTol, to $10^{-6}$ and $10^{-8}$, respectively. We checked that the solutions were not affected by a further decrease of RelTol and AbsTol.

The first set of ODEs involves Eqs. (1)-(8), (14), and (15). The second set involves matrix equation (43) with initial condition (44). To solve the matrix equation using ODE45, we reshaped the matrices into column vectors before transferring them to the derivative routine. Within the derivative routine, we reshaped the column vectors back into matrix form to calculate the derivatives and then transformed the output of the routine into column vector form to transfer it to the main program. The third set involves the mean age system (47) with initial condition (48).

To check that matrix B is calculated correctly, we solved Eq. (16) and compared the result with the solution of Eqs. (1)-(8). The same boundary conditions given by Eqs. (14) and (15) were utilized in both cases. The obtained solutions were identical (data not shown). To check that the matrix $\Phi\left(t, t_0\right)$ obtained by numerically solving Eqs. (43), (44) is correct, we compared the computed result with the result given by the following equation [18]:

$$\Phi\left(t, t_0\right) = \exp\left[\left(t - t_0\right)\mathrm{B}\right], \tag{S12}$$

which is valid for a real-valued constant square matrix B.

The numerical solution of Eqs. (47), (48) for the mean age of DCVs was validated by directly computing the mean age by integrating the age density distributions of DCVs, following the definition of the mean age given by Eq. (46).

## S5. Investigating sensitivity of the mean age of DCVs in boutons to the initial DCV concentration in the transiting state, $n_{0,t}$, and the portion of DCVs re-released from the resident state that join the anterograde pool, $\varepsilon$

In order to investigate the sensitivity of the solution to model parameters whose values are hardest to estimate, $n_{0,t}$ and $\varepsilon$, we also calculated the local sensitivity coefficients, which are first-order partial derivatives of the observables with respect to model parameters [28-31]. For example, the sensitivity coefficient of the mean age of resident DCVs in boutons to parameter $n_{0,t}$ at steady-state (ss) can be calculated as follows:

$$\frac{\partial \overline{a}_{i,ss}}{\partial n_{0,t}} \approx \left.\frac{\overline{a}_{i,ss}\left(n_{0,t} + \Delta n_{0,t}\right) - \overline{a}_{i,ss}\left(n_{0,t}\right)}{\Delta n_{0,t}}\right|_{\text{other parmeters kept constant}} \qquad (i=1,\ldots,26). \tag{S13}$$

where $\Delta n_{0,t} = 10^{-4} n_{0,t}$ (the accuracy was tested by using various step sizes).

The sensitivity coefficients were non-dimensionalized by introducing relative sensitivity coefficients [29,32], defined as (for example):

$$S_{n_{0,t}}^{\overline{\varepsilon}_{i,ss}} = \frac{n_{0,t}}{\overline{a}_{i,ss}} \frac{\partial \overline{a}_{i,ss}}{\partial n_{0,t}} \qquad (i=1,\ldots,26).$$ (S14)

The dimensionless sensitivity coefficients to $n_{0,t}$ and $\varepsilon$ are reported in Tables S5 and S6, respectively.

Table S5. Dimensionless sensitivity coefficients with respect to parameter $n_{0,t}$ for $\delta = 1$, $n_{0,t} = 1$, and $\varepsilon = 0.8$. Computations were performed with $\Delta n_{0,t} = 10^{-4} n_{0,t}$. An identical result was obtained for $\Delta n_{0,t} = 10^{-3} n_{0,t}$.

| $S_{n_{0,t}}^{\overline{a}_{1,ss}}$ | $S_{n_{0,t}}^{\overline{a}_{5,ss}}$ | $S_{n_{0,t}}^{\overline{a}_{13,ss}}$ | $S_{n_{0,t}}^{\overline{a}_{26,ss}}$ | $S_{n_{0,t}}^{\overline{a}_{1,t,ss}}$ | $S_{n_{0,t}}^{\overline{a}_{5,t,ss}}$ | $S_{n_{0,t}}^{\overline{a}_{13,t,ss}}$ | $S_{n_{0,t}}^{\overline{a}_{26,t,ss}}$ |
|---|---|---|---|---|---|---|---|
| 0.162 | 0.161 | 0.163 | 0.162 | 0.367 | 0.340 | 0.332 | 0.321 |

| $S_{n_{0,t}}^{\overline{a}_{2,tr,ss}}$ | $S_{n_{0,t}}^{\overline{a}_{5,tr,ss}}$ | $S_{n_{0,t}}^{\overline{a}_{13,tr,ss}}$ | $S_{n_{0,t}}^{\overline{a}_{26,tr,ss}}$ |
|---|---|---|---|
| 0.373 | 0.390 | 0.419 | 0.454 |

The sensitivity of the mean DCV age in the resident state in boutons to $n_{0,t}$ is positive in all boutons (Table S5). This is because a larger number of DCVs in the transiting state leads to more DCVs that have already spent some time in the resident state of more proximal boutons. This increases the average DCV age in boutons, which is consistent with the results reported in Table 2, compare line 4 ($n_{0,t} = 0.1$) with line 2 ($n_{0,t} = 1$) in Table 2.

Table S6. Dimensionless sensitivity coefficients with respect to parameter $\varepsilon$ for $n_{0,t} = 1$ and $\varepsilon = 0.8$. Computations were performed with $\Delta \varepsilon = 10^{-4} \varepsilon$. An identical result was obtained for $\Delta \varepsilon = 10^{-3} \varepsilon$.

| $S_\varepsilon^{\bar{a}_{1,ss}}$ | $S_\varepsilon^{\bar{a}_{5,ss}}$ | $S_\varepsilon^{\bar{a}_{13,ss}}$ | $S_\varepsilon^{\bar{a}_{26,ss}}$ | $S_\varepsilon^{\bar{a}_{1,r,ss}}$ | $S_\varepsilon^{\bar{a}_{5,ra,ss}}$ | $S_\varepsilon^{\bar{a}_{13,ra,ss}}$ | $S_\varepsilon^{\bar{a}_{26,ra,ss}}$ |
|---|---|---|---|---|---|---|---|
| 0.212 | 0.207 | 0.152 | 0.012 | 0.482 | 0.521 | 0.601 | 0.794 |

| $S_\varepsilon^{\bar{a}_{2,tr,ss}}$ | $S_\varepsilon^{\bar{a}_{5,tr,ss}}$ | $S_\varepsilon^{\bar{a}_{13,tr,ss}}$ | $S_\varepsilon^{\bar{a}_{26,tr,ss}}$ |
|---|---|---|---|
| 0.473 | 0.425 | 0.220 | 0.000 |

The sensitivity of the mean DCV age in the resident state in boutons to $\varepsilon$ is almost zero in the most proximal bouton. This is because most of the DCVs in the most proximal bouton are new as they enter from the axon. The sensitivity becomes positive in more distal boutons (Table S6). This is because the increase of the portion of DCVs released into the anterograde component makes the overall age of DCVs larger, since DCVs have a chance to be recaptured in the downstream boutons. This is also consistent with the results reported in Table 2 (compare the first three lines in Table 2).

## S6. Supplementary figures

### S6.1. Verifying the accuracy of computed age density distributions of DCVs in the resident and transiting states

In order to verify the accuracy of age density distributions of resident (Fig. 5a) and transiting (Figs. 5b and 5c) DCVs, we integrated the age densities over the age (from 0 to infinity). The integral should give the number of DCVs in the corresponding state (resident or transiting) of a bouton. The results indicate an excellent agreement between the value of the integral and the number of DCVs computed by numerically solving Eqs. (1)-(8) with boundary conditions (14), (15) (Fig. S1).

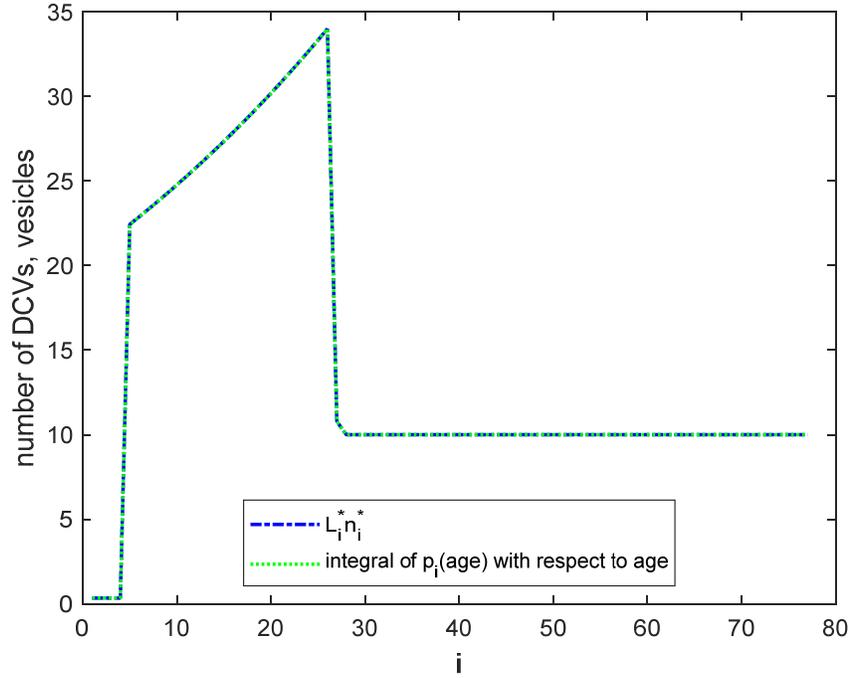

Fig. S1. Validation of the accuracy of computed age density of DCVs in the resident and transiting states (plotted in Fig. 5). Fig. S1 compares a value of $\int_0^\infty p_i(c,t)\,dc$ with a value of $L_i^* n_i^*$ for the steady-state situation. Here $i=1,...,52$; $n_1^* = n_1,...,n_{26}^* = n_{26}, n_{27}^* = n_{1,t},...,n_{52}^* = n_{26,t}$; and $L_1^* = L_1,...,L_{26}^* = L_{26}, L_{27}^* = L_1,...,L_{52}^* = L_{26}$. $i=1,...,26$ denotes the resident state in bouton $i$, $i=27$ denotes the transiting state in bouton 1, $i=28,...52$ denotes the anterograde transiting state in bouton $i-26$, and $i=53,...77$ denotes the retrograde transiting state in bouton $i-51$. $\delta=1$, $n_{0,t}=1$, and $\varepsilon=0.8$.

## S6.2. DCV age density distributions and mean DCV ages for the case when all captured DCVs are destroyed in boutons, $\delta=0$

For the case with no DCV re-release into the circulation ($\delta=0$), the age density distributions of transiting DCVs, displayed in Figs. 2b,c, lose their rightward skewed shape that they exhibited for $\delta=1$ (Figs. 5b,c). Notably, all the curves displayed in Fig. S2c are now symmetrical bell-shaped curves. Although the mean DCV age is smaller for $\delta=0$ (compare Fig. S3 with Fig. 6 and Table S4 with Table 2), the mode (peak location) is greater (compare Figs. S2b,c with Figs. 5b,c). The difference in the position of the peaks is explained by the fact that for $\delta = 1$ (Figs. 5b,c) the flux of transiting DCVs in the terminal is larger because DCVs that are re-released from the resident state join with DCVs that are coming from the axon. The flux of transiting DCVs flushes out older DCVs from the transiting state, which causes the peak to occur at a smaller age. For $\delta = 0$ captured DCVs are destroyed in boutons and the only source of transiting DCVs are those coming from the axon. Because for $\delta = 0$ the DCV flux is smaller, it takes longer time to flush out the older DCVs from the transiting state, which results in the peaks occurring at larger ages (Figs. S2b,c).

The smaller mean age of DCVs for $\delta = 0$ is due to the lack of the rightward skewed shape of the DCV age density distribution, which for $\delta = 1$ shifts the mean of the curve toward older ages (compare Table S4 with the second line in Table 2). The fact that for $\delta = 0$ the age of resident DCVs in proximal boutons exceeds the age of resident DCVs in distal boutons (Fig. S3a, Table S4), which contradicts experimental observations reported in [7], supports the model described by $\delta = 1$ (when DCVs captured into the resident state are re-released to the circulation).

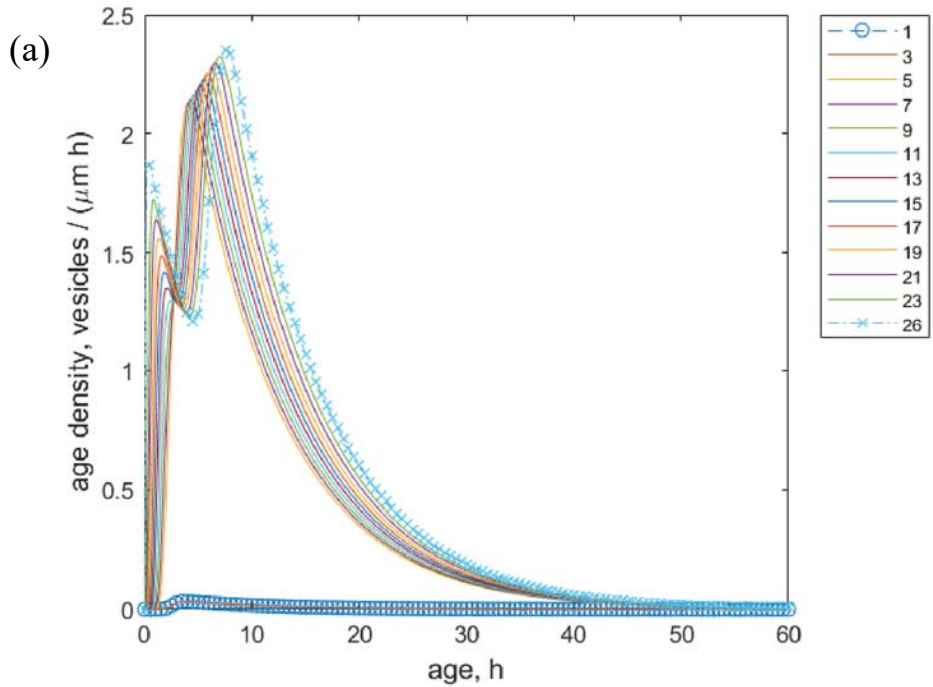

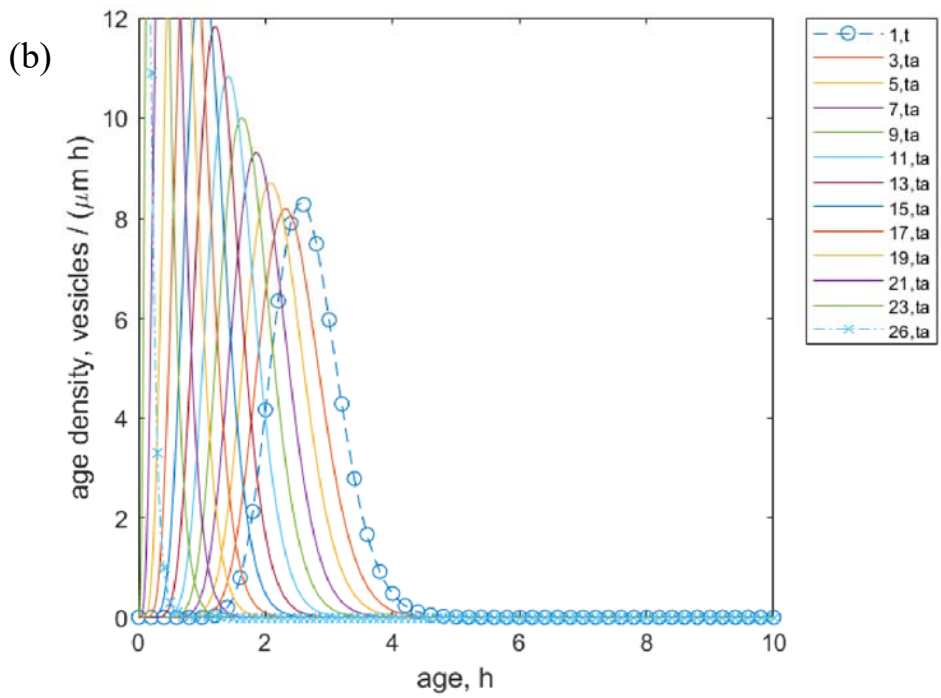

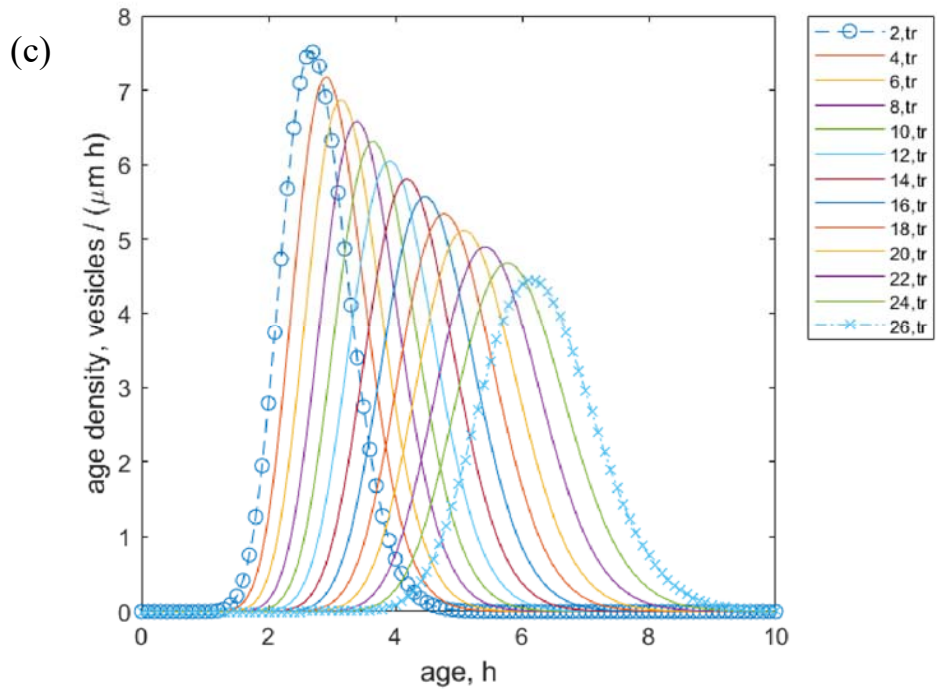

Fig. S2. (a) Age density of DCVs in the resident state in various boutons at steady-state. (b) Age density of DCVs in the anterograde transiting state in various boutons at steady-state. (c) Age density of DCVs in the retrograde transiting state in various boutons at steady-state. $\delta = 0$,

$n_{0,t} = 1$, and $\varepsilon = 0.8$. Age density is shown in every second bouton to make the figures less cluttered.

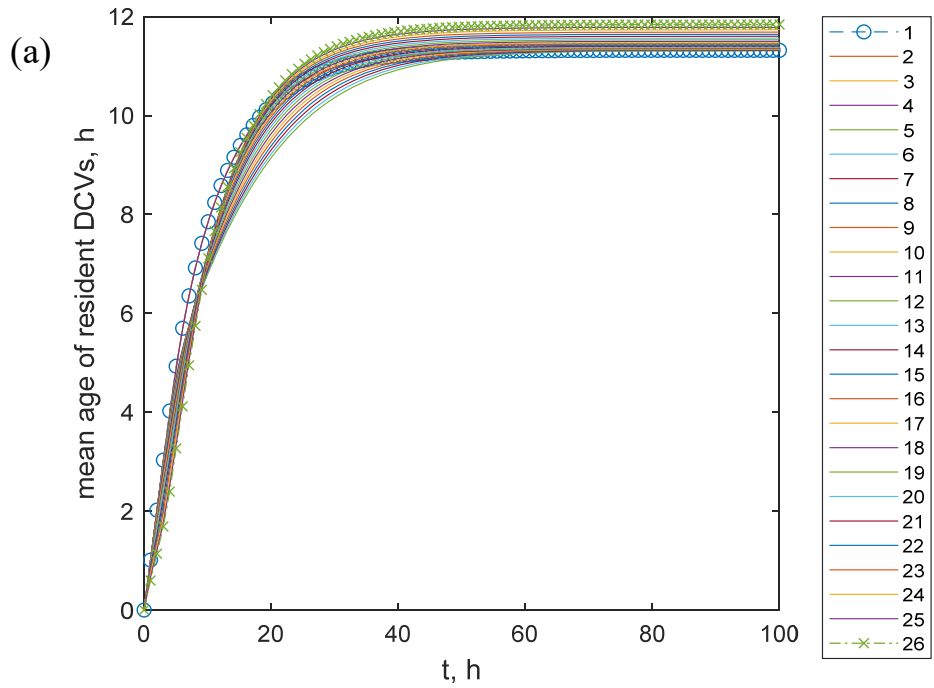

(a)

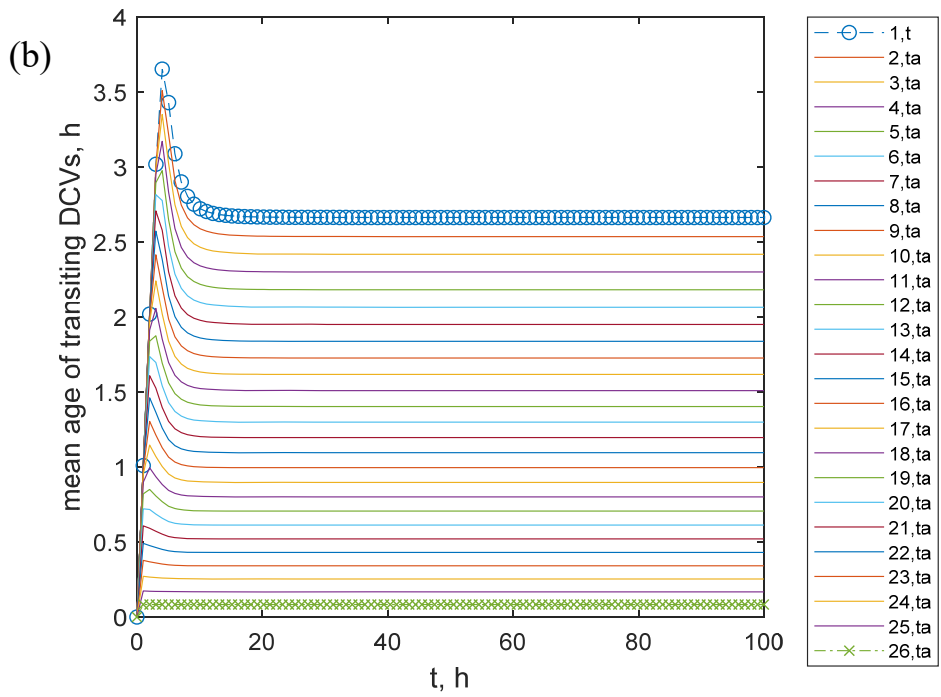

(b)

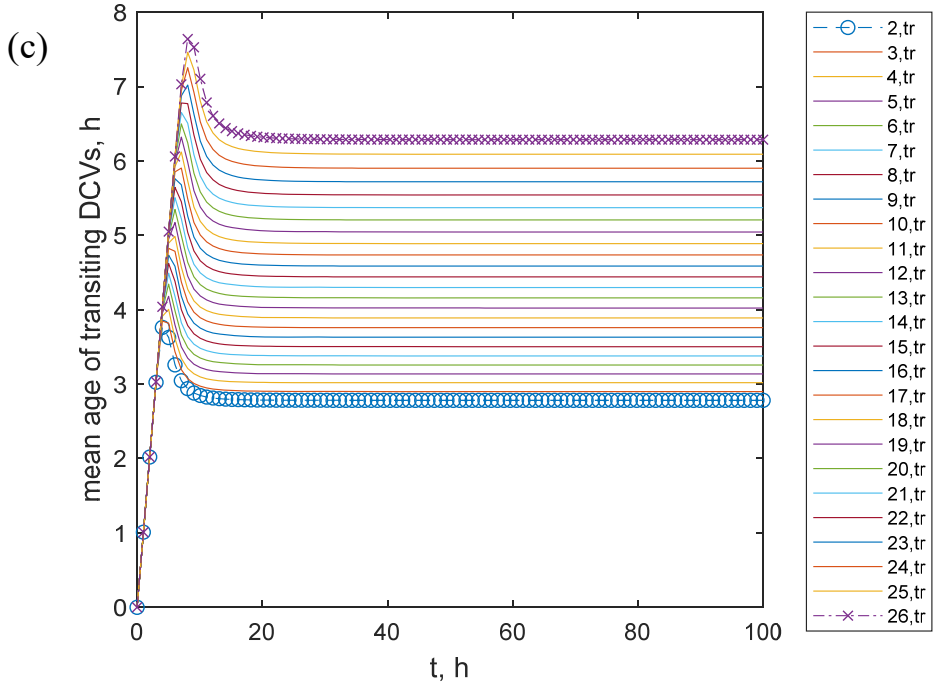

Fig. S3. (a) Mean age of resident DCVs in various boutons versus time. (b) Mean age of anterograde transiting DCVs in various boutons versus time. (c) Mean age of retrograde transiting DCVs in various boutons versus time. $\delta = 0$, $n_{0,t} = 1$, and $\varepsilon = 0.8$.

**S6.3. Effect of parameter $\varepsilon$ on anterograde and retrograde fluxes between the boutons**

If $\varepsilon = 1$, for a given time, anterograde fluxes increase from more proximal to more distal boutons, while retrograde fluxes decrease from more proximal to more distal boutons (Fig. 4a, S5a, S6a, and S7a). This happens because DCVs re-released from the resident state join the anterograde component of the DCV flux. For $\varepsilon = 0.8$, half of the re-released DCVs join the anterograde pool and half join the retrograde pool (Fig. S4b, S5b, S6b, and S7b). As a result, for a given time, anterograde and retrograde fluxes do not change significantly from bouton to bouton. For $\varepsilon = 0$, re-released DCVs join the retrograde component, therefore, for a given time, anterograde fluxes decrease from more proximal to more distal boutons, while retrograde fluxes increase from more proximal to more distal boutons (Fig. S4c, S5c, S6c, and S7c).

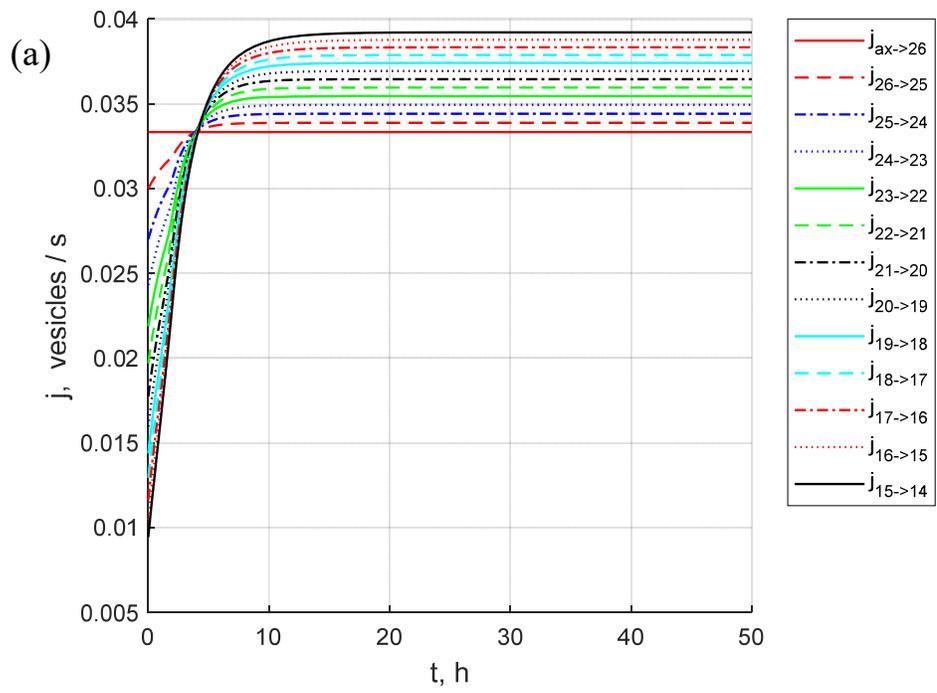

(a)

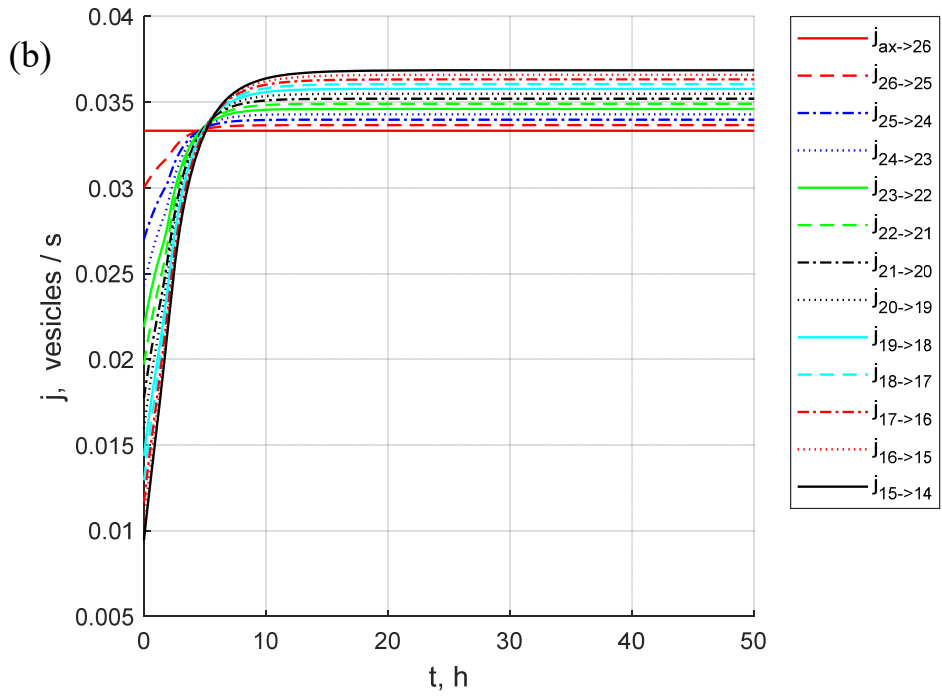

(b)

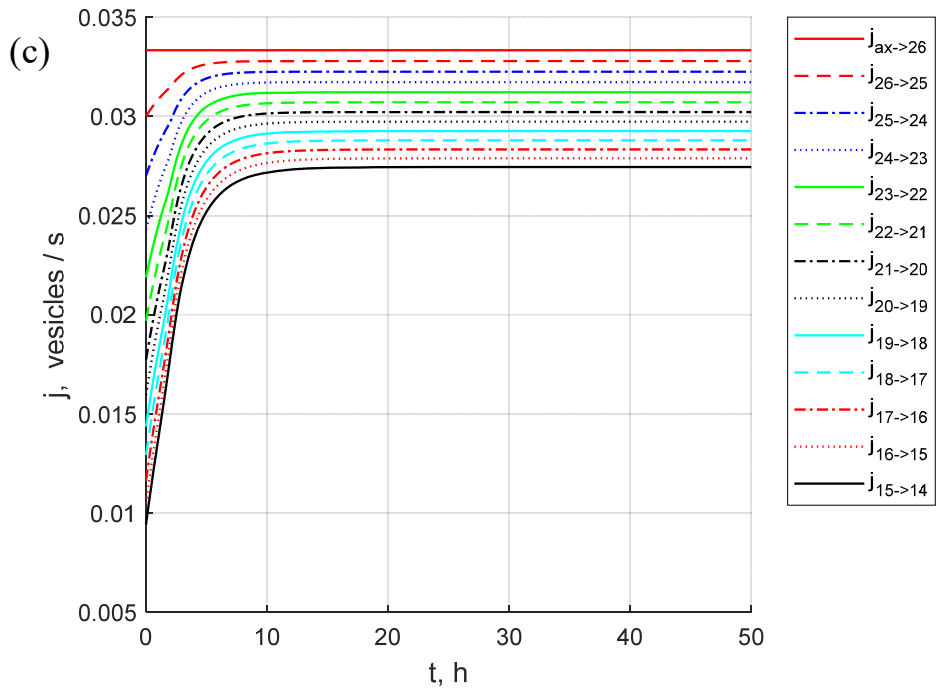

Fig. S4. The effect of parameter $\varepsilon$ on anterograde fluxes between the boutons. (a) $\varepsilon = 1$: all DCVs re-released from the resident state join the anterograde component. (b) $\varepsilon = 0.8$: 80% of DCVs that are re-released from the resident state join the anterograde component and 20% join the retrograde join the component. (c) $\varepsilon = 0$: all DCVs that are re-released from the resident state join the retrograde component. Fluxes ax→26 through 1→14. $\delta = 1$.

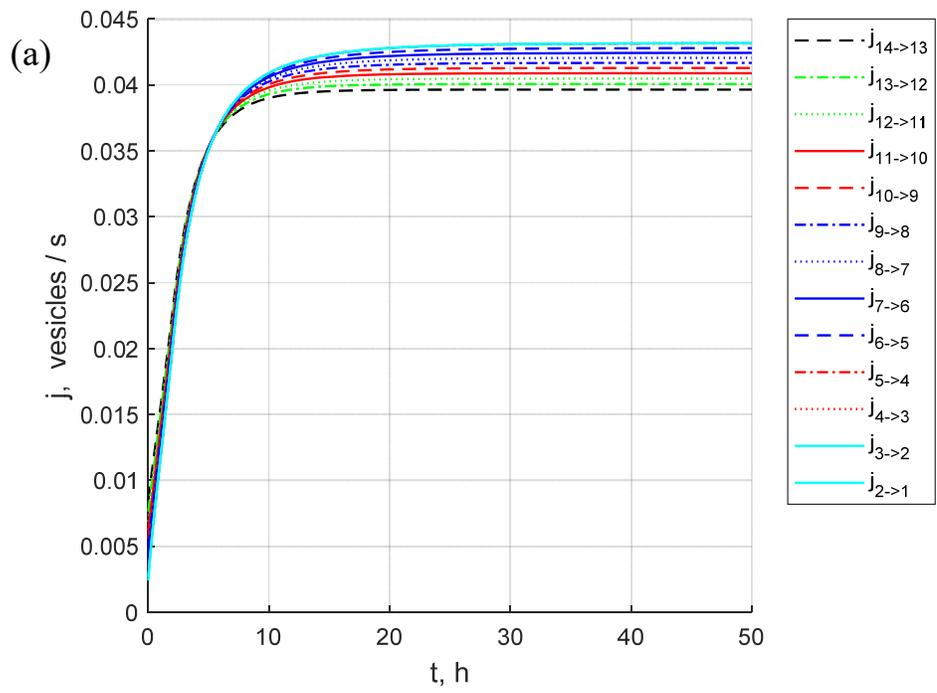

(a)

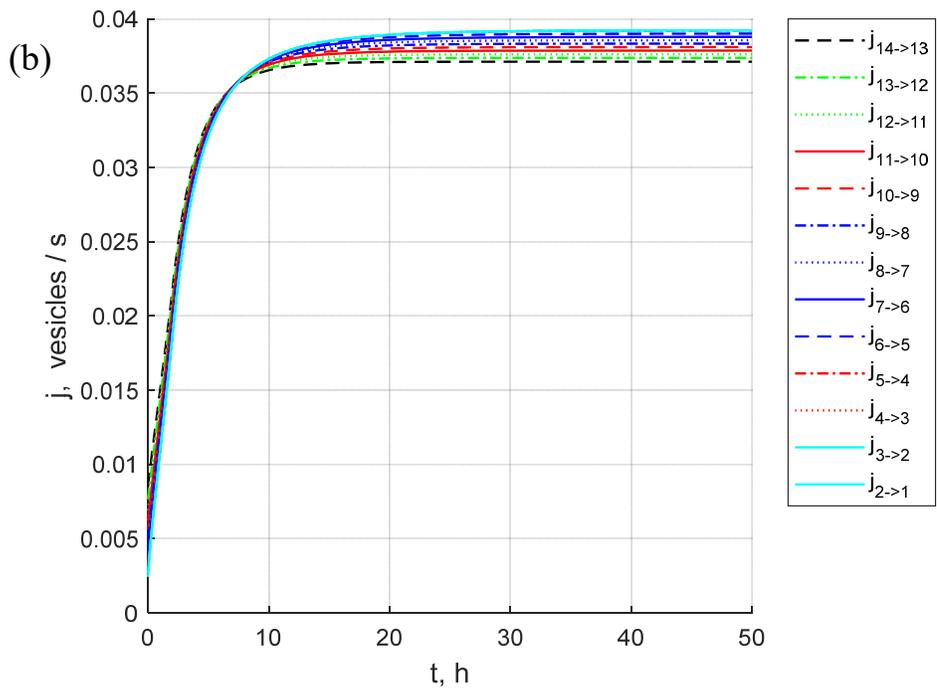

(b)

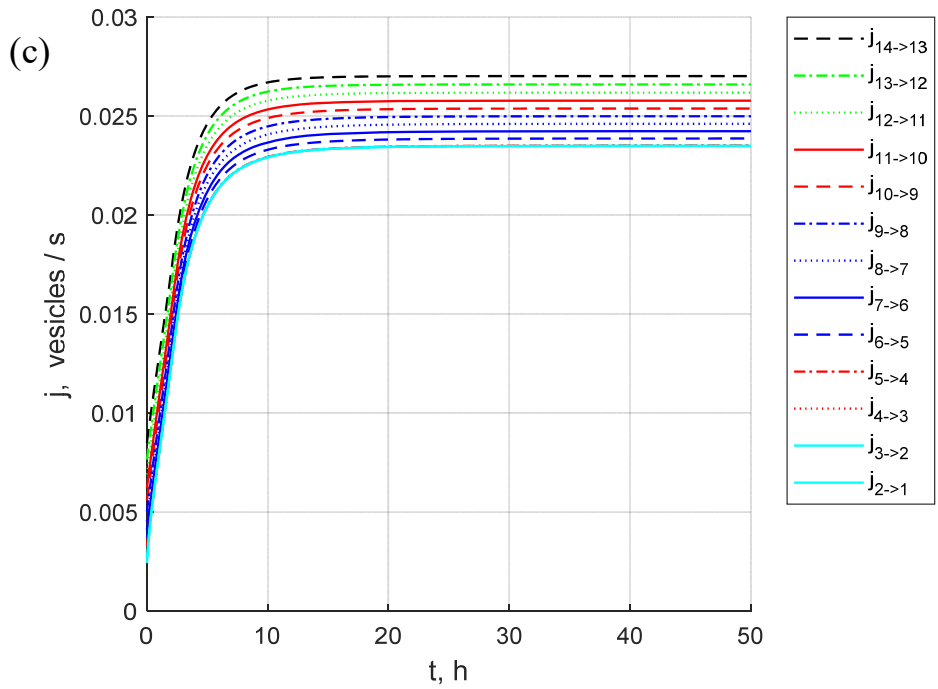

Fig. S5. The effect of parameter $\varepsilon$ on anterograde fluxes between the boutons. (a) $\varepsilon = 1$: all DCVs re-released from the resident state join the anterograde component. (b) $\varepsilon = 0.8$: 80% of DCVs that are re-released from the resident state join the anterograde component and 20% join the retrograde join the component. (c) $\varepsilon = 0$: all DCVs that are re-released from the resident state join the retrograde component. Fluxes 14→13 through 2→1. $\delta = 1$.

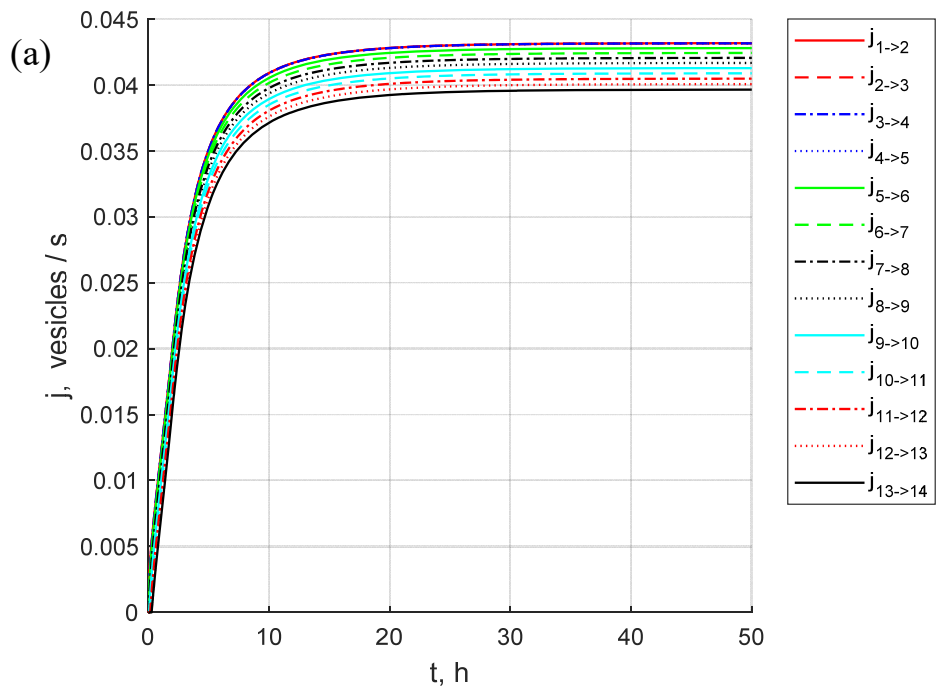

(a)

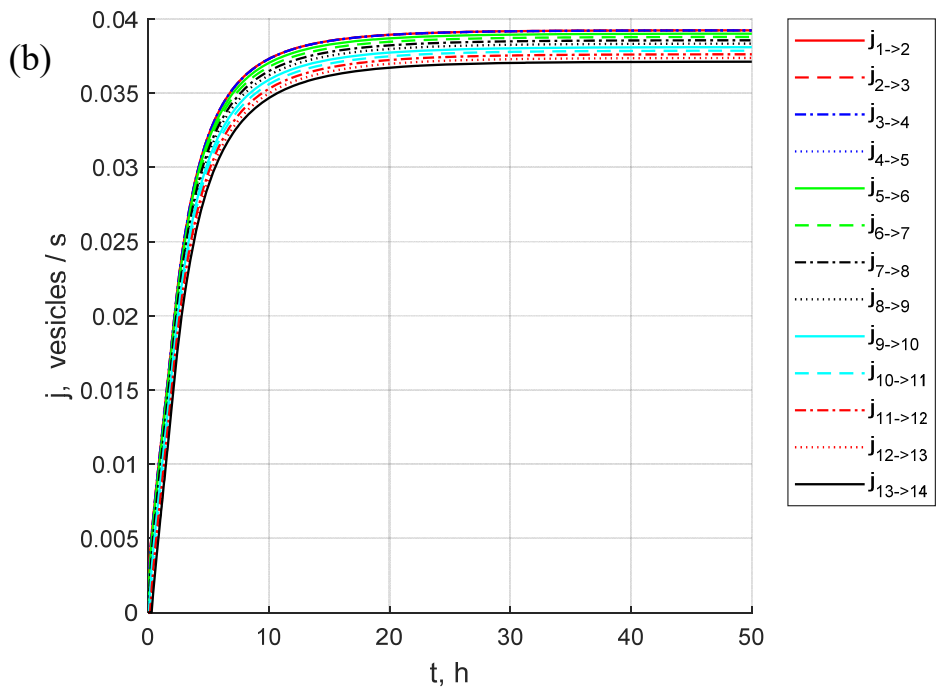

(b)

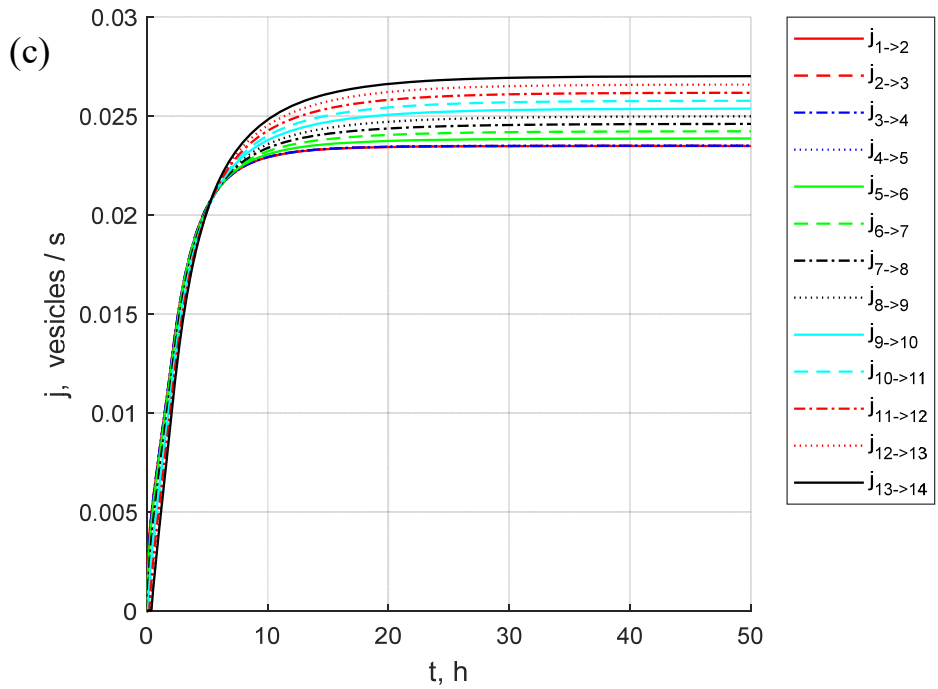

Fig. S6. The effect of parameter $\varepsilon$ on anterograde fluxes between the boutons. (a) $\varepsilon = 1$: all DCVs re-released from the resident state join the anterograde component. (b) $\varepsilon = 0.8$: 80% of DCVs that are re-released from the resident state join the anterograde component and 20% join the retrograde join the component. (c) $\varepsilon = 0$: all DCVs that are re-released from the resident state join the retrograde component. Fluxes 1→2 through 13→14. $\delta = 1$.

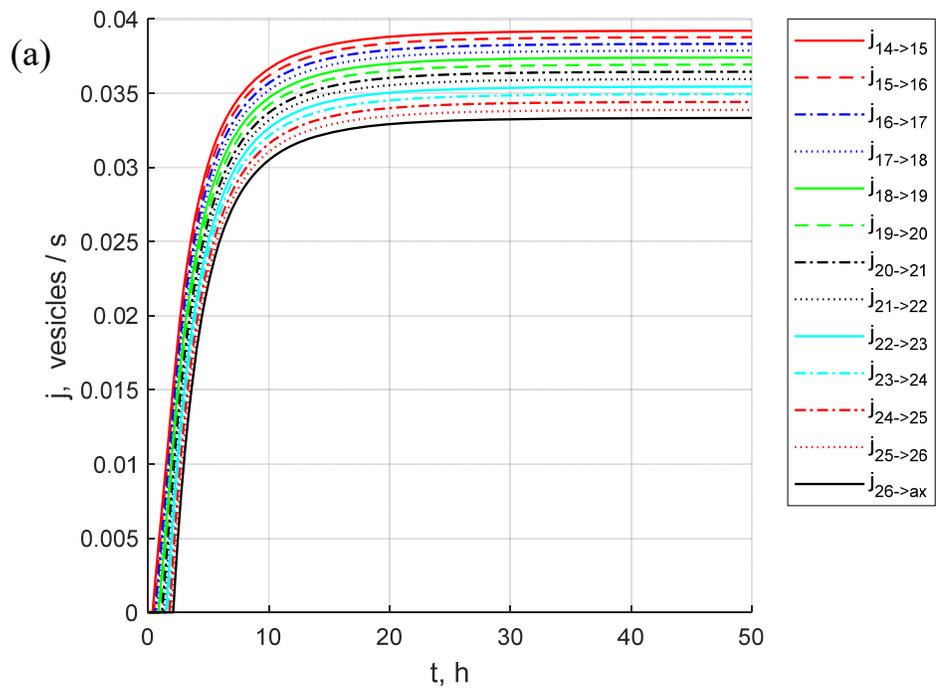

(a)

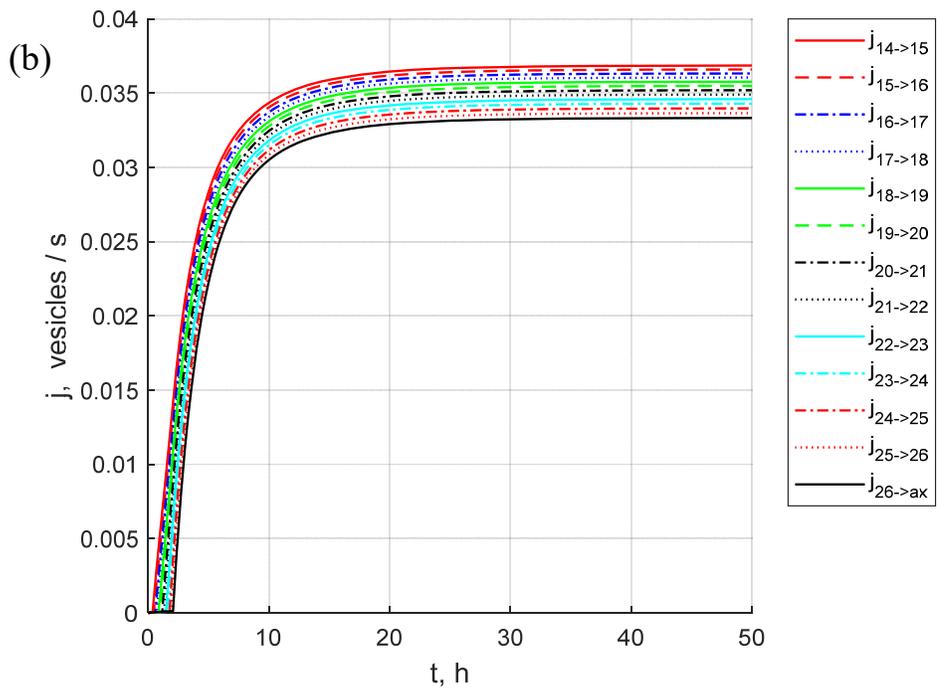

(b)

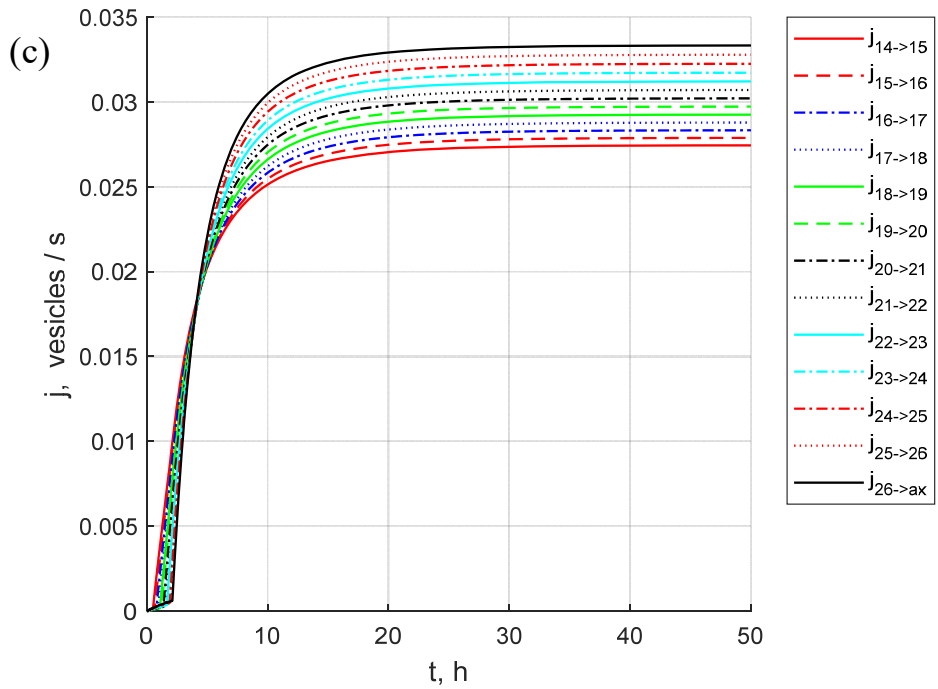

Fig. S7. The effect of parameter $\varepsilon$ on anterograde fluxes between the boutons. (a) $\varepsilon = 1$: all DCVs re-released from the resident state join the anterograde component. (b) $\varepsilon = 0.8$: 80% of DCVs that are re-released from the resident state join the anterograde component and 20% join the retrograde join the component. (c) $\varepsilon = 0$: all DCVs that are re-released from the resident state join the retrograde component. Fluxes 14→15 through 26→ax. $\delta = 1$.